\documentclass[twocolumn,secnumarabic,amssymb, nobibnotes, prl, reprint, superscriptaddress
]{revtex4-1}

\newcommand{\bbet}{\mathrm{B}_\mathrm{\beta}}
\newcommand{\bpar}{\mathrm{B}_\parallel}
\newcommand{\bper}{\mathrm{B}_\perp}
\newcommand{\bx}{\mathrm{B}_\mathrm{x}}
\newcommand{\by}{\mathrm{B}_\mathrm{y}}
\newcommand{\bz}{\mathrm{B}_\mathrm{z}}
\newcommand{\rnl}{\mathrm{R}_\mathrm{NL}}
\newcommand{\rsq}{\mathrm{R}_\mathrm{sq}}
\newcommand{\vf}{\mathrm{v}_\mathrm{F}}
\newcommand{\dsp}{\mathrm{D}_{\mathrm{s}\parallel}}
\newcommand{\ds}{\mathrm{D}_{\mathrm{s}}}
\newcommand{\dc}{\mathrm{D}_{\mathrm{c}}}
\newcommand{\tsp}{\tau_{\mathrm{s}\parallel}}

\newcommand{\ohm}{\Omega}
\newcommand{\tpar}{\tau_\parallel}
\newcommand{\tper}{\tau_\perp}
\newcommand{\tiv}{\tau_\mathrm{iv}}
\newcommand{\tp}{\tau_\mathrm{p}}
\newcommand{\tbet}{\tau_\beta}

\usepackage{graphicx}
\usepackage{color}

\usepackage{amsmath}
\usepackage{dcolumn}
\usepackage{bm}
\usepackage{float}
\usepackage{ulem}

\begin{document}

\title{Observation of spin-valley coupling induced large spin lifetime anisotropy in bilayer graphene}%

\author{Johannes Christian Leutenantsmeyer}
\email[]{These authors contributed equally to this work, \\E-Mail: j.c.leutenantsmeyer@rug.nl}
\affiliation{Physics of Nanodevices, Zernike Institute for Advanced Materials, University of Groningen, 9747 AG Groningen, The Netherlands}

\author{Josep Ingla-Ayn\'{e}s$^*$}
\affiliation{Physics of Nanodevices, Zernike Institute for Advanced Materials, University of Groningen, 9747 AG Groningen, The Netherlands}





\author{Jaroslav Fabian}
\affiliation{Institute for Theoretical Physics, University of Regensburg, 93040 Regensburg, Germany}

\author{Bart J. van Wees}
\affiliation{Physics of Nanodevices, Zernike Institute for Advanced Materials, University of Groningen, 9747 AG Groningen, The Netherlands}

\begin{abstract}
We report the first observation of a large spin lifetime anisotropy in bilayer graphene (BLG) fully encapsulated between hexagonal boron nitride. We characterize the out-of-plane ($\tper$) and in-plane ($\tpar$) spin lifetimes by oblique Hanle spin precession. At 75~K and the charge neutrality point (CNP) we observe a strong  anisotropy of $\tau_\perp/\tau_\parallel$ = 8 $\pm$ 2. This value is comparable to graphene/TMD heterostructures, whereas our high quality BLG provides with $\tper$ up to 9~ns, a more than two orders of magnitude larger spin lifetime. The anisotropy decreases to 3.5 $\pm$ 1 at a carrier density of n = $6\times 10^{11}~$cm$^{-2}$. 
Temperature dependent measurements show above 75~K a decrease of $\tau_\perp/\tau_\parallel$ with increasing temperature, reaching the isotropic case close to room temperature. We explain our findings with electric field induced spin-valley coupling arising from the small intrinsic spin orbit fields in BLG of 12~$\mu$eV at the CNP.

\end{abstract}

\date{\today}%
\maketitle

Coupling between the electronic spin and valley degree of freedom arises in materials without inversion symmetry such as single layer transition metal dichalcogenides (TMDs) \cite{Xiao2012,Schaibley2016} where the electronic bands are spin split by the spin-orbit fields. 
Due to time reversal symmetry, the induced spin splitting is opposite for the K and K' points of the Brillouin zone. 
This leads to a coupling between the spin and valley degrees of freedom, and enables new functionalities such as the optical injection of spin currents with circularly polarized light \cite{Luo2017,Avsar2017a}. The spin-valley coupling has been imprinted on the band structure of monolayer graphene by placing it in proximity with a TMD and measured using spin \cite{Cummings2017,Ghiasi2017,Benitez2017} and charge transport \cite{Wang2015a,Wang2016a,Zihlmann2018}. However, it remains a question if similar behavior can be observed in pristine graphene devices.

BLG has an intrinsic spin-orbit coupling (SOC) of $\lambda_\mathrm{I} \sim 12\,\mu$eV, which points out of the BLG plane. A perpendicular electric field breaks the inversion symmetry and, as a consequence, the intrinsic SOC induces an out-of-plane spin splitting of $2 \lambda_\mathrm{I} \sim$ 24~$\mu$eV at the K points \cite{Konschuh2012}. The splitting has opposite sign in K and K' and therefore a valley dependence. Recent ab-initio calculations show that the encapsulation of BLG in hexagonal boron nitride (hBN) preserves the presence of the spin splitting with a similar magnitude \cite{Fabiannew}.

Thermal broadening and inhomogeneities due to doping fluctuations \cite{Martin2008} prevent the direct measurement of such a small spin splitting in conventional charge transport experiments. However, spin precession experiments can resolve spin splittings much smaller than k$_\mathrm{B}$T, if the splitting extends over a sufficiently large region in reciprocal space and energy \cite{Leutenantsmeyer2017}. In the presence of an out-of-plane spin splitting, the dephasing of spins follows the Dyakonov-Perel mechanism \cite{VanTuan2016}. The in-plane spin lifetime $\tpar$ is inversely proportional to the intervalley scattering time, $\tpar \propto \lambda_\mathrm{I}^2/\tiv$ \cite{Cummings2017}. 
Hence, $\tpar$ is sensitive to the SOC strength. 

\begin{figure}[!tb]
\centerline{\includegraphics[width=0.9\linewidth]{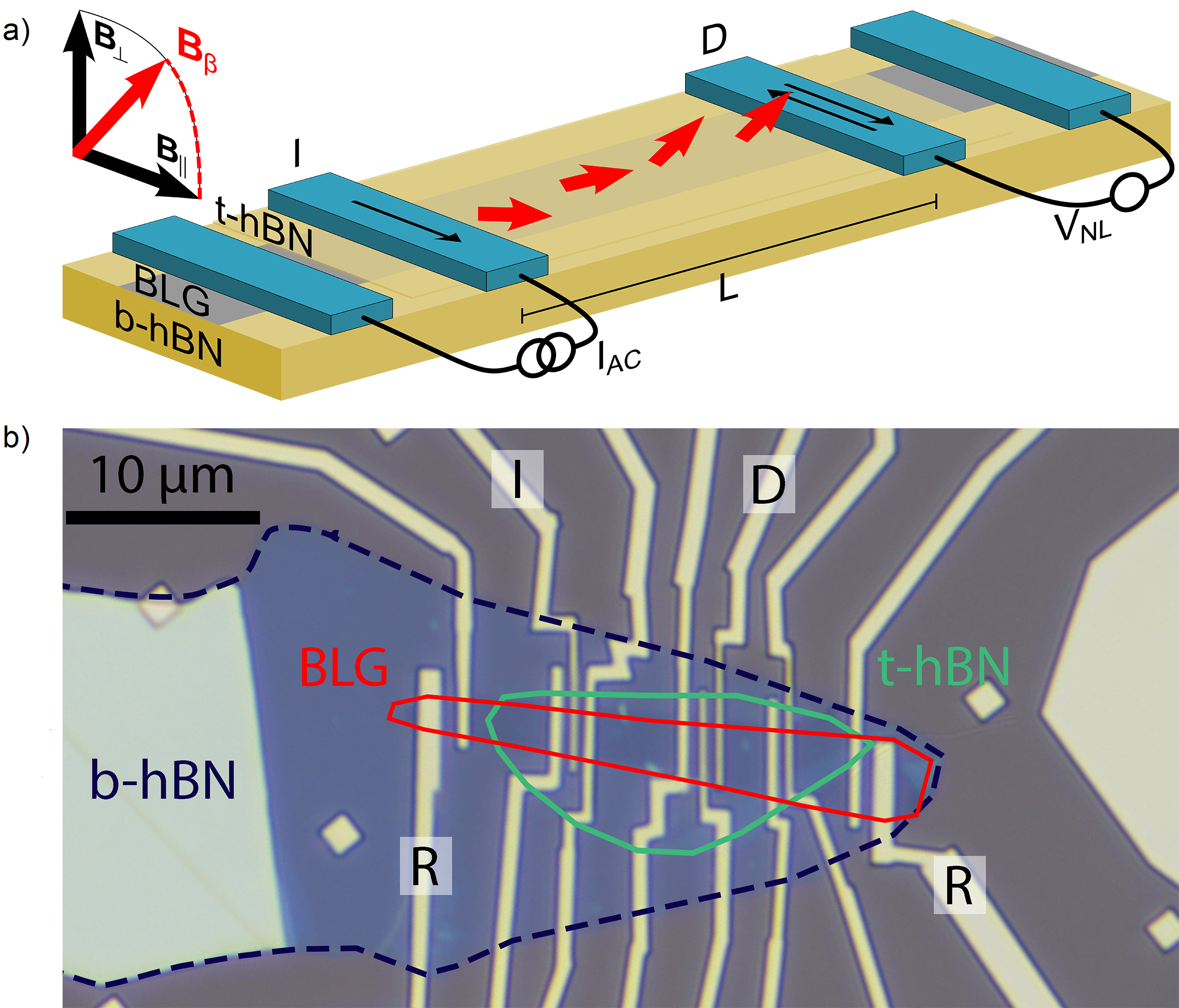}}
\caption{Schematic a) and optical image b) of the device geometry. BLG is encapsulated by a 1~nm thick hBN tunnel barrier (t-hBN) and a 5~nm (b-hBN) flake. A low frequency AC current (I$_\mathrm{AC}$) injects a spin accumulation into the BLG. The non local signal (V$_\mathrm{NL}$) is measured using standard lock-in technique. The precession of injected in-plane spins around the magnetic field $\bbet$ is illustrated in the encapsulated BLG channel. Note that the outer reference contacts (R) are not covered by the hBN tunnel barrier. The injector (I) and detector contact (D) used for the measurements discussed in the main text are labeled and have a spacing of L = 7~$\mu$m. 
\label{Figure1}}
\end{figure}


\begin{figure*}[!t]
\centerline{\includegraphics[width=1\linewidth]{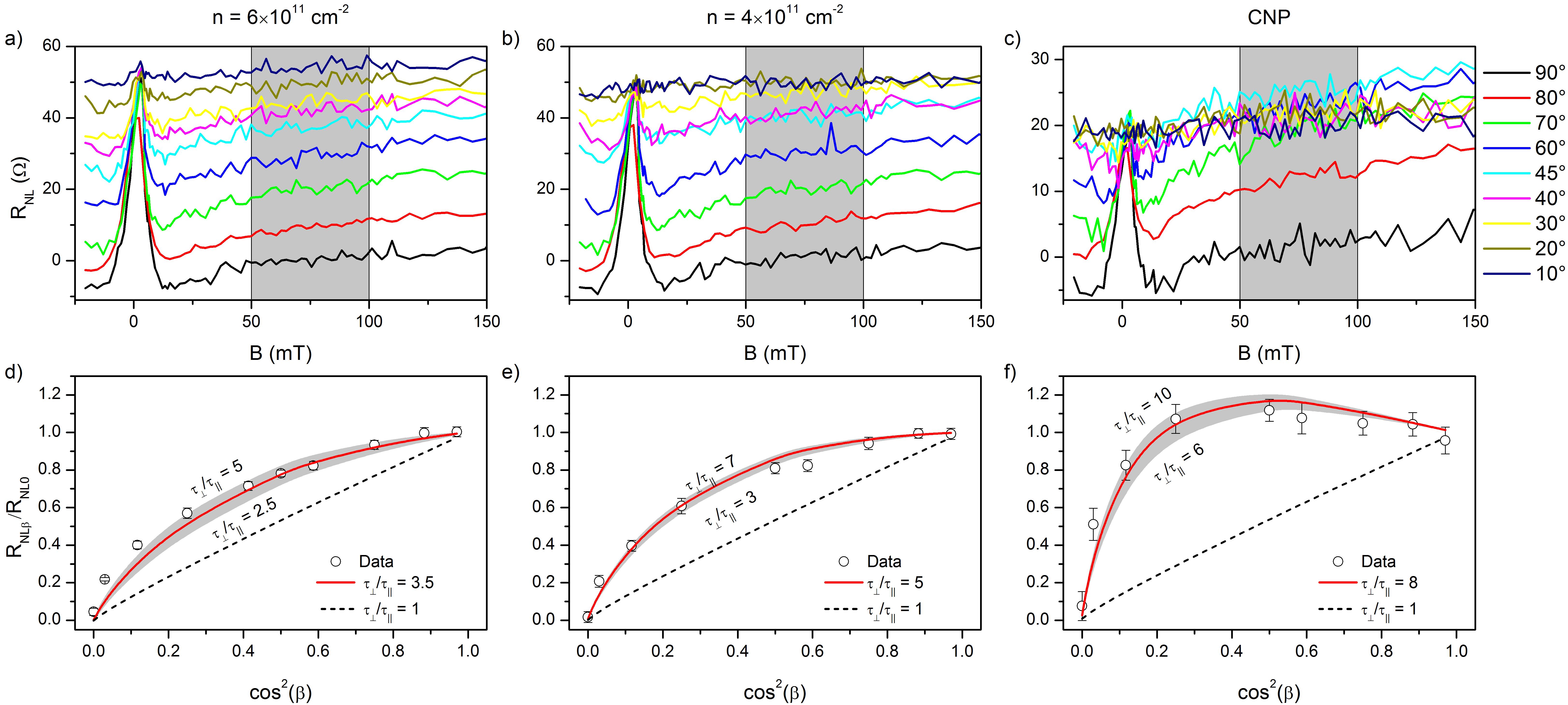}}
\caption{Oblique Hanle spin precession data for a) n = $6\times 10^{11}~$cm$^{-2}$, b) n = $4\times 10^{11}~$cm$^{-2}$ and c) the CNP. $\mathrm{R}_{\mathrm{NL}0}$ denotes the non local resistance at zero field and $\mathrm{R}_{\mathrm{NL}\beta}$ the non local resistance where the perpendicular spin component has fully dephased. $\mathrm{R}_{\mathrm{NL}\beta}$ is obtained by averaging $\rnl$ over the shaded area (50 - 100~mT). The bottom panels d - e show the comparison between the the ratios $\mathrm{R}_{\mathrm{NL}\beta}/\mathrm{R}_\mathrm{NL0}$ and our model for different anisotropy values. The shaded area corresponds to the estimated error margin with the denoted anisotropy values. Note that panels a-c have a small background in $\rnl$ of 9.3~$\ohm$, 18~$\ohm$ and 17.8~$\ohm$ subtracted. 
\label{Figure4}}
\end{figure*}

Apart from the intrinsic SOC, breaking of the inversion symmetry leads to Rashba spin orbit fields in the graphene plane \cite{Rashba2009,Guimaraes2014} that affect both in-plane and out-of-plane ($\tper$) spin lifetimes. Therefore, spin relaxation in BLG is a result of an interplay between between intrinsic and Rashba SOC. The Rashba SOC depends on the Fermi velocity, which increases with the carrier density n, whereas the intrinsic spin orbit splitting decreases with n. As a consequence, the spin lifetime anisotropy ($\tper/\tpar$) is expected to depend strongly on n near the CNP \cite{Konschuh2012,Wang2013} allowing the electrical control of the spin lifetime anisotropy.



Here we study $\tper$ and $\tpar$ in fully hBN encapsulated BLG using oblique spin precession. Our results show that, in contrast with monolayer graphene \cite{Tombros2008, Guimaraes2014, Raes2016,Ringer2017}, at temperatures below 300~K, the ratio $\tper/\tpar$ is significantly above 1 over the full measured range of n. 
At 75~K we observe a dependence of $\tper/\tpar$ on the carrier concentration which increases from 3.5 $\pm$ 1 at n = $6\times 10^{11}~$cm$^{-2}$ to 8 $\pm$ 2 at the CNP confirming the role of the spin-valley coupling on the spin transport. The anisotropy at the CNP is comparable to graphene/TMD systems \cite{Ghiasi2017,Benitez2017}. However, the spin lifetimes in our BLG devices are two orders of magnitude larger \cite{Han2011,Yang2011,Avsar2011,Neumann2013a,Ingla-Aynes2015,Avsar2016}. These results show that small spin orbit fields can induce sizable effects on the spin relaxation and indicate that the spin relaxation in our devices is limited by $\lambda_\mathrm{I}$ and Rashba SOC.

\begin{figure*}[tbh]
\centerline{\includegraphics[width=1\linewidth]{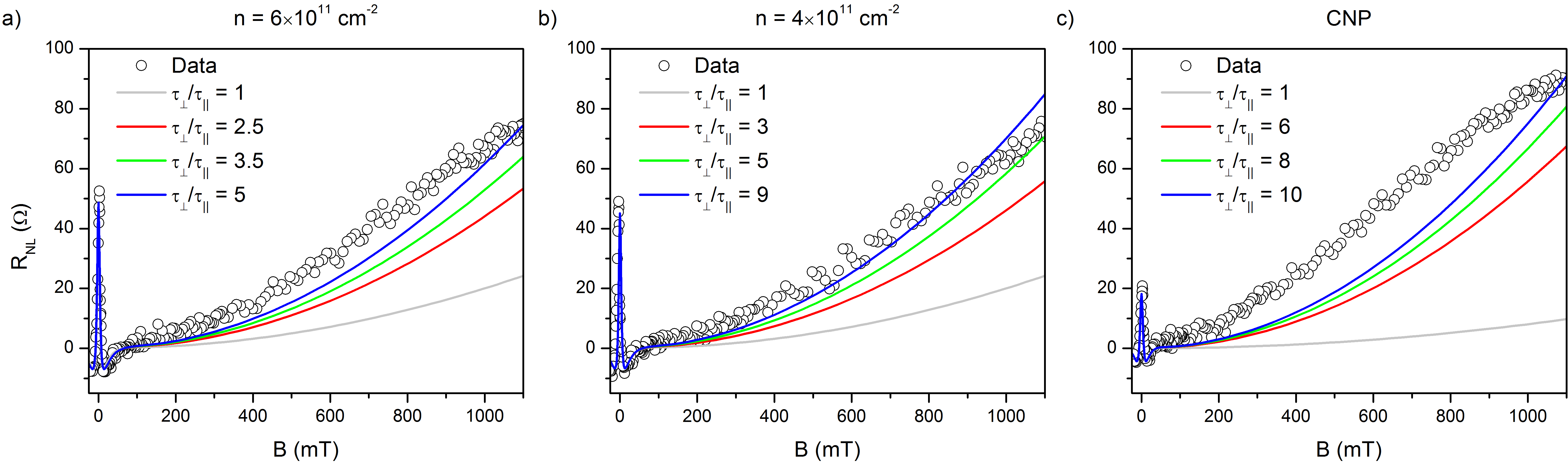}}
\caption{High field Hanle spin precession curves at $\beta = 90^\circ$ and T = 75~K for the three discussed carrier concentrations. We simulate the spin precession using the parameters from Fig.~\ref{Figure4}. The gray line corresponds to the isotropic case. The perpendicular saturation field of the cobalt contacts is 1.5~T. Note that the same background as in Fig.~\ref{Figure4} has been subtracted. 
\label{Figure5}}
\end{figure*}

The device is shown in Fig.~\ref{Figure1} where the BLG is protected from contamination by a trilayer hBN tunnel barrier on top and a 5~nm thick bottom hBN flake below \cite{Zomer2014}. The stack is deposited on a 90~nm SiO$_2$/Si wafer which is used as a backgate. Ferromagnetic cobalt contacts are defined using standard e-beam lithography and e-beam evaporation techniques and are used for spin injection and detection. With a back gate we tune the carrier concentration from the hole regime, slightly beyond the CNP ($2\times 10^{11}~$cm$^{-2}$) up to $6\times 10^{11}~$cm$^{-2}$ in the electron regime. The CNP is at -2~V applied to the backgate, indicating a small background doping. The electric field at the CNP is estimated ti be between 10 and 20~mV/nm \footnote{See Supplemental Material.}. Note that the application of large electric fields (above 2~V/nm) to BLG can result in bandgaps up to 200~meV \cite{Castro2007,Oostinga2008,Zhang2009}. However, the small fields applied to our sample lead to bandgap openings significantly smaller than k$_\mathrm{B}$T and are neglected in our analysis.

The mobility $\mu$ of the sample is 12000~cm$^{2}$/Vs at n = $4\times 10^{11}~$cm$^{-2}$ obtained using $\mu$ = 1/e d$\sigma$/dn where $\sigma$ is the conductivity and e the electron charge. The charge diffusion coefficient is $\mathrm{D}_\mathrm{c}$ = 0.026~m$^2$/s, which is in agreement with the spin diffusion coefficient $\mathrm{D}_\mathrm{s}$ = (0.021 $\pm$ 0.005)~m$^2$/s obtained from Hanle spin precession. This indicates the consistency of the analysis. 

To optimize the spin injection efficiency, we apply additionally to the AC measurement current a DC bias current of -0.6~$\mu$A to the trilayer hBN barrier \cite{Gurram2017,Gurram2018}. Note that the negative bias applied to the injector causes a sign change in the spin polarization of the injector and therefore in $\rnl$. For comparison with conventional Hanle curves, we have inverted the sign of $\rnl$ (see \cite{Note1}).

Fig.~\ref{Figure4}a-c shows the experimental results obtained from oblique Hanle spin precession experiments (see Fig.~\ref{Figure1}a) for the schematics of the measurement) at three different carrier densities. The data shown in panels a and d is measured at n = $6\times 10^{11}~$cm$^{-2}$, b and e at n = $4\times 10^{11}~$cm$^{-2}$, whereas the data in c and f is measured at the CNP. $\mathrm{R}_{\mathrm{NL}\beta}$ is defined as the spin signal where the spin accumulation perpendicular to the magnetic field $\bbet$ is fully dephased. We extract $\mathrm{R}_{\mathrm{NL}\beta}$ from the experiment by averaging $\rnl$ between 50 and 100~mT, indicated by the gray area at low magnetic fields in Fig.~\ref{Figure4}a-c.

The spins are injected collinear to the in-plane magnetization of the ferromagnetic electrode with efficiency P. Since only the component parallel to $\bbet$ is conserved, the injection and detection efficiencies for the measured spins become P$\times\cos(\beta)$. Consequently, $\mathrm{R}_{\mathrm{NL}\beta}$ is proportional to $\cos^2(\beta)$. Therefore, at $\beta = 45^\circ$, one would expect $\mathrm{R}_{\mathrm{NL}\beta}$ to be reduced by 50\% compared to $\mathrm{R}_{\mathrm{NL}0}$ in an isotropic system. We find at 75~K that at all different carrier concentrations in Fig.~\ref{Figure4}a-c, $\mathrm{R}_{\mathrm{NL}\beta}/\mathrm{R}_{\mathrm{NL}0}$ is clearly above 0.5 for $\beta = 45^\circ$, which can only be the case if $\tper/\tpar>1$. This can be seen from Eq.~\ref{SLA}, which can be used to quantify the degree of anisotropy \cite{Raes2016,Raes2017}:
\begin{align}
\frac{\mathrm{R}_{\mathrm{NL}\beta}}{\mathrm{R}_\mathrm{NL0}} &= \sqrt{\frac{\tbet}{\tpar}}\exp{\left[ \frac{-\mathrm{L}}{\lambda_\parallel} \left( \sqrt{\frac{\tpar}{\tbet}} -1\right) \right]}\cos^2{(\beta)} \label{SLA}\\
\frac{\tbet}{\tpar} &= \left( \cos^2(\beta)+ \frac{\tpar}{\tper}\sin^2(\beta)\right)^{-1}
\end{align}
However, this model is only applicable for a channel significantly longer than both in-plane and out-of-plane spin relaxation length.  
The out-of-plane spin relaxation length ($\sim$12~$\mu$m) is longer than the closest spacing between sample edge and the injector (8~$\mu$m). Therefore, the exact device geometry has to be taken into account for a quantitative analysis. 

To carefully account for the device geometry, we solve the Bloch equations for anisotropic spin transport numerically. Furthermore, we include both the effect of $\bbet$ on the contact magnetization direction using a Stoner-Wohlfarth model and the influence of the finite resistances of the reference contacts \cite{Note1,Maassen2012,Zhu2018}. The Hanle precession curves are simulated for different ratios $\tper/\tpar$ and different angles $\beta$. We obtain $\mathrm{R}_{\mathrm{NL}\beta}/\mathrm{R}_\mathrm{NL0}$ from the simulated curves using the same procedure as for the experimental data. 

The resulting curves are shown in Fig.~\ref{Figure4}d to f where the red solid line represents the best fit to the data. The gray areas correspond to the estimated error margin with the annotated values. The case of an isotropic system is shown by the dotted gray lines. We find $\tper/\tpar$ to be $3.5 \pm 1$ (n = $6\times 10^{11}~$cm$^{-2}$), $5 \pm 2$ (n = $4\times 10^{11}~$cm$^{-2}$), and $8 \pm 2$ (CNP). We have measured and analyzed different contact spacings and different injector/detector contact pairs which all showed a consistent behavior and are discussed in the supplementary information \cite{Note1}. 

When a large $\bper$ is applied, the Co magnetization direction rotates out of the sample plane. As a consequence, a perpendicular spin component is injected making $\rnl$ sensitive to the spin lifetime anisotropy \cite{Tombros2008}. 
The data measured up to a large $\bper$ is shown in Fig.~\ref{Figure5} together with the simulated Hanle curves. It should be noted that for all carrier concentrations $\rnl$($\bper$ = 1.1~T) clearly exceeds $\rnl$($\bper$ = 0~T), which is a direct consequence of $\tper > \tpar$.
The Hanle curves are simulated for different $\tper/\tpar$ ratios, where the gray lines represent the isotropic case. We attribute the difference between the low (Fig.~\ref{Figure4}) and high field analysis (Fig.~\ref{Figure5}) to two origins. Firstly, our simulations use a simple out-of-plane shape anisotropy model to describe the rotation of the electrode magnetizations under $\bper$ whereas the magnetization behavior can deviate from the idealized system. Secondly, we observe magnetoresistance of the BLG channel, which can reach up to 50\% at high fields and at the CNP. Its possible influence on the measured data is discussed in the supplementary information \cite{Note1}. However, for magnetic fields below 0.1~T at the CNP the magnetoresistance is below 1\%. Hence, magnetoresistance does not affect our low field analysis. 

\begin{figure}[tb]
\centerline{\includegraphics[width=1\linewidth]{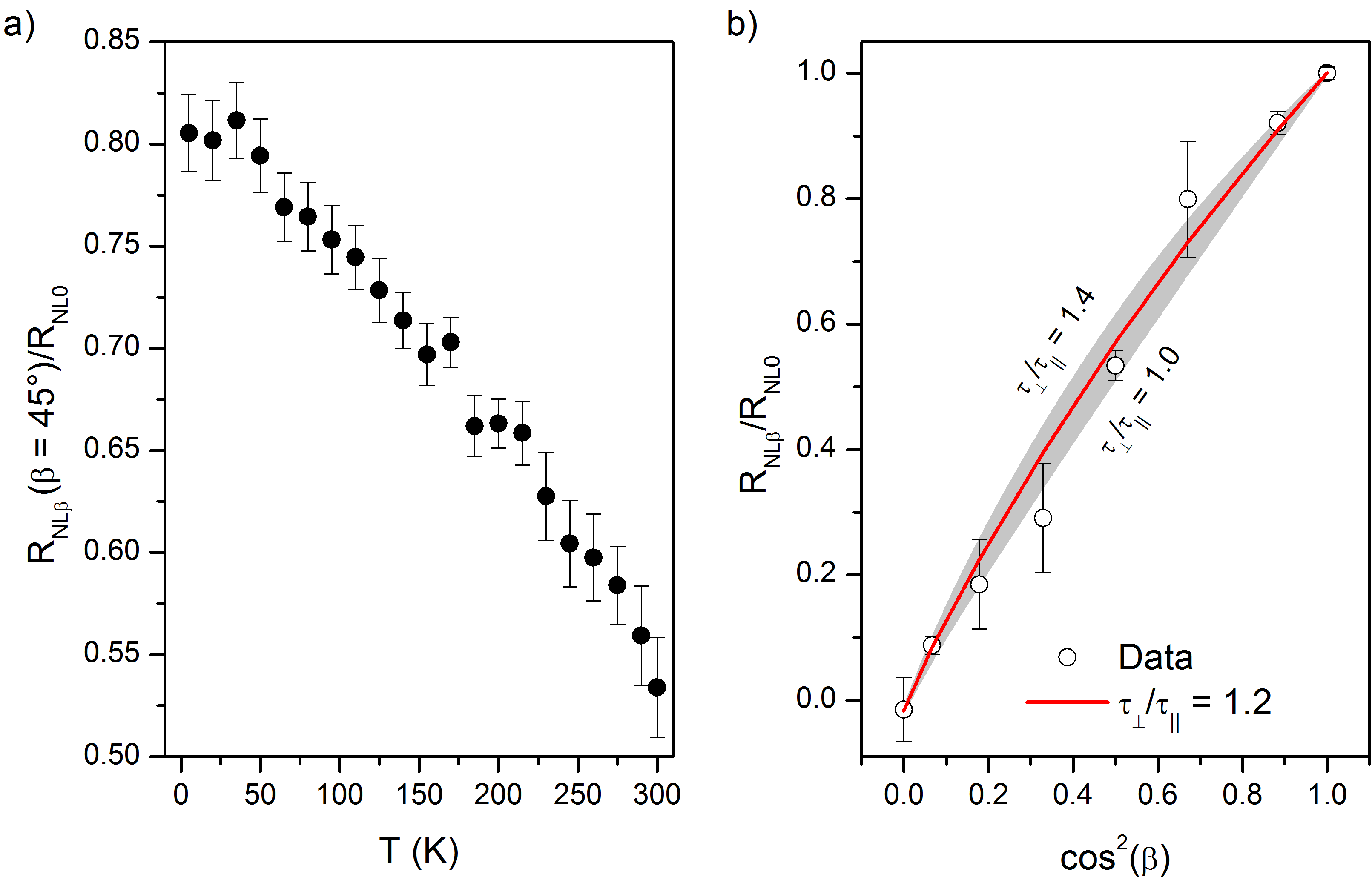}}
\caption{(a) Temperature dependence of the ratio $\mathrm{R}_{\mathrm{NL}\beta}/\mathrm{R}_\mathrm{NL0}$ measured at $\beta = 45^\circ$. The trend towards $\mathrm{R}_{\mathrm{NL}\beta}/\mathrm{R}_\mathrm{NL0} = 0.5$ with increasing temperature implies that the anisotropy decreases. b) Extraction of the $\tper/\tpar$ for T = 300~K analogous to Fig.~\ref{Figure4}. We conclude that $\tper \sim \tpar$ at room temperature. 
\label{Figure6}}
\end{figure}

We can estimate the intervalley scattering time $\tiv$ from the extracted $\tpar$ and $\tper$ by assuming a Dyakonov-Perel type of spin relaxation as predicted theoretically \cite{Cummings2017,VanTuan2016}:
\begin{align}
\frac{1}{2\tper} + \left(\frac{2\lambda_\mathrm{I}}{\hbar}\right)^2\tiv = \frac{1}{\tpar} \label{tiveqn}
\end{align}
where $1/\tper = (2\lambda_\mathrm{R}/\hbar)^2\tp$ with the Rashba SOC $\lambda_\mathrm{R}$. The relevant spin and charge transport parameters are shown in table~\ref{table1main}. We observe the shortest $\tiv$ at the CNP, which we attribute to two origins: Firstly, $\lambda_\mathrm{I}$ is 12~$\mu$eV at the CNP but decays quickly with increasing momentum from the CNP \cite{Konschuh2012}. As a consequence, the effective $\lambda_\mathrm{I}$ is smaller than 12~$\mu$eV and our extracted $\tiv$ should be seen as lower bound. Secondly, the spin splittings have opposite sign in the conduction and valence bands. Hence, non energy conserving scattering between both bands plays the same role as intervalley scattering when both electrons and holes contribute to the transport. $\tiv$ becomes an effective parameter ($\tiv^*$) determined by both intervalley and interband scatting ($\tau_\mathrm{ib}$), $\tiv^{*-1} = \tau_\mathrm{ib}^{-1}+\tiv^{-1}$.

\begin{table}[bt]
\caption{Spin and charge transport parameters of the discussed device. $\tiv$ is calculated using Eq.~\ref{tiveqn}. The density dependence of $\lambda_\mathrm{I}$ is extracted from \cite{Konschuh2012}. The momentum scattering time $\tp$ is obtained assuming $\ds = \dc = \vf^2\tp/2$, where $\vf$ is the Fermi velocity.}
\begin{ruledtabular}
\begin{tabular}{c c c c c c c c c}
\renewcommand{\arraystretch}{2}
T & n & $\rsq$ & $\ds$ & $\tpar$ & $\tper$ & $\lambda_\mathrm{I}$ & $\tiv$ & $\tp$ \\
K &  cm$^{-2}$& $\ohm$& $\mathrm{m^2/s}$ & ns & ns & $\mu$eV & ps  & ps \\ \hline
75 & CNP & 1550 & 0.010 & 1.1 & 8.8 & 12 &0.6& -\\
75&4$\times$10$^{11}$& 900 & 0.018 & 1.9 & 9.4 &2& 12 &0.28 \\
75&6$\times$10$^{11}$& 750 & 0.021 & 1.7 & 6.1 & 1 &45 & 0.22\\
300 & 4$\times$10$^{11}$ & 510 & 0.03 & 1.2 & 1.4 & 2 & 4& 0.40
\end{tabular}
\end{ruledtabular}
\label{table1main}
\end{table}
Note that the values of $\lambda_\mathrm{I}$ from table~\ref{table1main} are calculated in pristine BLG with an applied electric field of 25~mV/nm \cite{Konschuh2012}. The accurate determination of $\lambda_\mathrm{I}$ from first principles requires the knowledge of the alignment between the crystal planes of hBN and BLG. However, preliminary ab-initio calculations support the presence of a spin splitting in the range of 24~$\mu$eV at the K and K' points in hBN encapsulated BLG under small electric fields \cite{Fabiannew}.

It should be mentioned that our out-of-plane spin lifetimes in BLG (up to 9~ns) are close to the largest measured lifetimes of 12~ns in SLG \cite{Drogeler2016}. Therefore, the spin relaxation length becomes comparable to the device size and uncertainties, such as the spin lifetime in the adjacent uncovered BLG regions, can affect the analysis. Moreover, it is not clear whether the spin relaxation follows purely the Dyakonov-Perel mechanism and if other sources of spin orbit coupling become relevant for limiting $\tpar$ and $\tper$ in BLG \cite{Kochan2015,Tuan2014,Han2014}. 

Lastly, we discuss the temperature dependence of the spin lifetime anisotropy. The carrier density dependence of $\tper/\tpar$ at T = 5~K is discussed in the supplementary information \cite{Note1} and gives comparable results to T = 75~K ($\tper/\tpar = 2$ at $6\times 10^{12}~$cm$^{-2}$ and $\tper/\tpar = 8$ at the CNP). 
Fig.~\ref{Figure6}a shows the ratio $\mathrm{R}_{\mathrm{NL}\beta}/\mathrm{R}_\mathrm{NL0}$ measured at an angle of $\beta = 45^\circ$ and zero backgate voltage (n = $4\times 10^{11}~$cm$^{-2}$, measured at 5~K and 75~K). 
We observe a continuous decrease of $\mathrm{R}_{\mathrm{NL}\beta}/\mathrm{R}_\mathrm{NL0}$ as the temperature increases. At room temperature $\mathrm{R}_{\mathrm{NL}\beta}/\mathrm{R}_\mathrm{NL0}$ is close to 0.5, which corresponds to an isotropic system where $\tper/\tpar \approx 1$. 
The full angular dependence of $\mathrm{R}_{\mathrm{NL}\beta}/\mathrm{R}_\mathrm{NL0}$ at T = 300~K is shown in Fig.~\ref{Figure6}b. We extract here $\tper/\tpar$ = 1.2, where we estimate the error margin to be between 1 and 1.4. Due to an increased gate leakage current, we are unable to reach the CNP at 300~K. Therefore, we assume that the doping of the BLG flake remains constant over the measured temperature range and consequently the carrier concentration at room temperature is $4\times 10^{11}~$cm$^{-2}$. We calculate $\tp \approx $ 0.4~ps $\sim \tiv/10$ indicating that the decrease of anisotropy at 300~K is caused by the decrease of $\tiv$. Note that the thermal broadening at 300~K causes a sizable spread in momenta that can lead to lower lifetime anisotropies because $\lambda_\mathrm{I}$ diminishes fast with increasing n. 

Theoretical calculations predict in contrast to our results a maximum of the anisotropy around 175~K \cite{Wang2013}. Additionally, the anisotropy is predicted to be below 1 at low temperatures due to the suppression of intervalley scattering induced by electron-phonon interaction. Both predictions are not consistent with our observations, which we attribute to two main differences between theory and experiment. Firstly, the calculations are performed at n = $3\times 10^{12}~$cm$^{-2}$, which is significantly above n for our device. As we have demonstrated in this letter, the anisotropy is strongly affected by n. Secondly, our device is fully encapsulated in hBN, which can affect the phonon modes in BLG. 
At room temperature, these calculations predict $\tper/\tpar$ above 50 with $\tpar$ greater than 10~ns, whereas we find an almost isotropic system and $\tpar$ = 1.2~ns. 

In summary, we have studied the spin lifetime anisotropy in BLG by oblique spin precession. $\tper$ is found to be up to 8 times larger than $\tpar$ at the CNP. The anisotropy is found to decrease with increasing carrier concentration. An increase in temperature above 75~K causes a decrease of $\tper/\tpar$ and around room temperature $\tper$ approaches a similar value as $\tpar$, implying that BLG becomes isotropic. We attribute this to the intrinsic out-of-plane spin orbit fields in BLG, which, despite of their small magnitude, induce a significant spin-valley coupling that can be used to control spins in BLG \cite{Konschuh2012,Wang2013}.

The authors acknowledge fruitful discussions with M.~Gmitra, A.A.~Kaverzin and K.~Zollner. This project has received funding from the European Union’s Horizon 2020 research and innovation program under grant agreement No 696656 (‘Graphene Flagship’), the Marie Curie initial training network ‘Spinograph’ (grant agreement No 607904), the DFG SFB~1277 (projects A09 and B07) and the Spinoza Prize awarded to B.J. van Wees by the ‘Netherlands Organization for Scientific Research’ (NWO).

\bibliography{references}

\begin{thebibliography}{43}%
\makeatletter
\providecommand \@ifxundefined [1]{%
 \@ifx{#1\undefined}
}%
\providecommand \@ifnum [1]{%
 \ifnum #1\expandafter \@firstoftwo
 \else \expandafter \@secondoftwo
 \fi
}%
\providecommand \@ifx [1]{%
 \ifx #1\expandafter \@firstoftwo
 \else \expandafter \@secondoftwo
 \fi
}%
\providecommand \natexlab [1]{#1}%
\providecommand \enquote  [1]{``#1''}%
\providecommand \bibnamefont  [1]{#1}%
\providecommand \bibfnamefont [1]{#1}%
\providecommand \citenamefont [1]{#1}%
\providecommand \href@noop [0]{\@secondoftwo}%
\providecommand \href [0]{\begingroup \@sanitize@url \@href}%
\providecommand \@href[1]{\@@startlink{#1}\@@href}%
\providecommand \@@href[1]{\endgroup#1\@@endlink}%
\providecommand \@sanitize@url [0]{\catcode `\\12\catcode `\$12\catcode
  `\&12\catcode `\#12\catcode `\^12\catcode `\_12\catcode `\%12\relax}%
\providecommand \@@startlink[1]{}%
\providecommand \@@endlink[0]{}%
\providecommand \url  [0]{\begingroup\@sanitize@url \@url }%
\providecommand \@url [1]{\endgroup\@href {#1}{\urlprefix }}%
\providecommand \urlprefix  [0]{URL }%
\providecommand \Eprint [0]{\href }%
\providecommand \doibase [0]{http://dx.doi.org/}%
\providecommand \selectlanguage [0]{\@gobble}%
\providecommand \bibinfo  [0]{\@secondoftwo}%
\providecommand \bibfield  [0]{\@secondoftwo}%
\providecommand \translation [1]{[#1]}%
\providecommand \BibitemOpen [0]{}%
\providecommand \bibitemStop [0]{}%
\providecommand \bibitemNoStop [0]{.\EOS\space}%
\providecommand \EOS [0]{\spacefactor3000\relax}%
\providecommand \BibitemShut  [1]{\csname bibitem#1\endcsname}%
\let\auto@bib@innerbib\@empty
\bibitem [{\citenamefont {Xiao}\ \emph {et~al.}(2012)\citenamefont {Xiao},
  \citenamefont {Liu}, \citenamefont {Feng}, \citenamefont {Xu},\ and\
  \citenamefont {Yao}}]{Xiao2012}%
  \BibitemOpen
  \bibfield  {author} {\bibinfo {author} {\bibfnamefont {D.}~\bibnamefont
  {Xiao}}, \bibinfo {author} {\bibfnamefont {G.-B.}\ \bibnamefont {Liu}},
  \bibinfo {author} {\bibfnamefont {W.}~\bibnamefont {Feng}}, \bibinfo {author}
  {\bibfnamefont {X.}~\bibnamefont {Xu}}, \ and\ \bibinfo {author}
  {\bibfnamefont {W.}~\bibnamefont {Yao}},\ }\href {\doibase
  10.1103/PhysRevLett.108.196802} {\bibfield  {journal} {\bibinfo  {journal}
  {Physical Review Letters}\ }\textbf {\bibinfo {volume} {108}},\ \bibinfo
  {pages} {1} (\bibinfo {year} {2012})}\BibitemShut {NoStop}%
\bibitem [{\citenamefont {Schaibley}\ \emph {et~al.}(2016)\citenamefont
  {Schaibley}, \citenamefont {Yu}, \citenamefont {Clark}, \citenamefont
  {Rivera}, \citenamefont {Ross}, \citenamefont {Seyler}, \citenamefont {Yao},\
  and\ \citenamefont {Xu}}]{Schaibley2016}%
  \BibitemOpen
  \bibfield  {author} {\bibinfo {author} {\bibfnamefont {J.~R.}\ \bibnamefont
  {Schaibley}}, \bibinfo {author} {\bibfnamefont {H.}~\bibnamefont {Yu}},
  \bibinfo {author} {\bibfnamefont {G.}~\bibnamefont {Clark}}, \bibinfo
  {author} {\bibfnamefont {P.}~\bibnamefont {Rivera}}, \bibinfo {author}
  {\bibfnamefont {J.~S.}\ \bibnamefont {Ross}}, \bibinfo {author}
  {\bibfnamefont {K.~L.}\ \bibnamefont {Seyler}}, \bibinfo {author}
  {\bibfnamefont {W.}~\bibnamefont {Yao}}, \ and\ \bibinfo {author}
  {\bibfnamefont {X.}~\bibnamefont {Xu}},\ }\href {\doibase
  10.1038/natrevmats.2016.55} {\bibfield  {journal} {\bibinfo  {journal}
  {Nature Reviews Materials}\ }\textbf {\bibinfo {volume} {1}},\ \bibinfo
  {pages} {16055} (\bibinfo {year} {2016})}\BibitemShut {NoStop}%
\bibitem [{\citenamefont {Luo}\ \emph {et~al.}(2017)\citenamefont {Luo},
  \citenamefont {Xu}, \citenamefont {Zhu}, \citenamefont {Wu}, \citenamefont
  {Joan}, \citenamefont {Zhan}, \citenamefont {Neupane},\ and\ \citenamefont
  {Kawakami}}]{Luo2017}%
  \BibitemOpen
  \bibfield  {author} {\bibinfo {author} {\bibfnamefont {Y.~K.}\ \bibnamefont
  {Luo}}, \bibinfo {author} {\bibfnamefont {J.}~\bibnamefont {Xu}}, \bibinfo
  {author} {\bibfnamefont {T.}~\bibnamefont {Zhu}}, \bibinfo {author}
  {\bibfnamefont {G.}~\bibnamefont {Wu}}, \bibinfo {author} {\bibfnamefont
  {E.}~\bibnamefont {Joan}}, \bibinfo {author} {\bibfnamefont {W.}~\bibnamefont
  {Zhan}}, \bibinfo {author} {\bibfnamefont {M.~R.}\ \bibnamefont {Neupane}}, \
  and\ \bibinfo {author} {\bibfnamefont {R.~K.}\ \bibnamefont {Kawakami}},\
  }\href {\doibase 10.1021/acs.nanolett.7b01393} {\bibfield  {journal}
  {\bibinfo  {journal} {Nano Letters}\ }\textbf {\bibinfo {volume} {17}},\
  \bibinfo {pages} {3877} (\bibinfo {year} {2017})}\BibitemShut {NoStop}%
\bibitem [{\citenamefont {Avsar}\ \emph {et~al.}(2017)\citenamefont {Avsar},
  \citenamefont {Unuchek}, \citenamefont {Liu}, \citenamefont {{Lopez
  Sanchez}}, \citenamefont {Watanabe}, \citenamefont {Taniguchi}, \citenamefont
  {{\"{O}}zyilmaz},\ and\ \citenamefont {Kis}}]{Avsar2017a}%
  \BibitemOpen
  \bibfield  {author} {\bibinfo {author} {\bibfnamefont {A.}~\bibnamefont
  {Avsar}}, \bibinfo {author} {\bibfnamefont {D.}~\bibnamefont {Unuchek}},
  \bibinfo {author} {\bibfnamefont {J.}~\bibnamefont {Liu}}, \bibinfo {author}
  {\bibfnamefont {O.}~\bibnamefont {{Lopez Sanchez}}}, \bibinfo {author}
  {\bibfnamefont {K.}~\bibnamefont {Watanabe}}, \bibinfo {author}
  {\bibfnamefont {T.}~\bibnamefont {Taniguchi}}, \bibinfo {author}
  {\bibfnamefont {B.}~\bibnamefont {{\"{O}}zyilmaz}}, \ and\ \bibinfo {author}
  {\bibfnamefont {A.}~\bibnamefont {Kis}},\ }\href {\doibase
  10.1021/acsnano.7b06800} {\bibfield  {journal} {\bibinfo  {journal} {ACS
  Nano}\ }\textbf {\bibinfo {volume} {11}},\ \bibinfo {pages} {11678} (\bibinfo
  {year} {2017})}\BibitemShut {NoStop}%
\bibitem [{\citenamefont {Cummings}\ \emph {et~al.}(2017)\citenamefont
  {Cummings}, \citenamefont {Garc{\'{i}}a}, \citenamefont {Fabian},\ and\
  \citenamefont {Roche}}]{Cummings2017}%
  \BibitemOpen
  \bibfield  {author} {\bibinfo {author} {\bibfnamefont {A.~W.}\ \bibnamefont
  {Cummings}}, \bibinfo {author} {\bibfnamefont {J.~H.}\ \bibnamefont
  {Garc{\'{i}}a}}, \bibinfo {author} {\bibfnamefont {J.}~\bibnamefont
  {Fabian}}, \ and\ \bibinfo {author} {\bibfnamefont {S.}~\bibnamefont
  {Roche}},\ }\href {\doibase 10.1103/PhysRevLett.119.206601} {\bibfield
  {journal} {\bibinfo  {journal} {Physical Review Letters}\ }\textbf {\bibinfo
  {volume} {119}},\ \bibinfo {pages} {206601} (\bibinfo {year}
  {2017})}\BibitemShut {NoStop}%
\bibitem [{\citenamefont {Ghiasi}\ \emph {et~al.}(2017)\citenamefont {Ghiasi},
  \citenamefont {Ingla-Ayn{\'{e}}s}, \citenamefont {Kaverzin},\ and\
  \citenamefont {{van Wees}}}]{Ghiasi2017}%
  \BibitemOpen
  \bibfield  {author} {\bibinfo {author} {\bibfnamefont {T.~S.}\ \bibnamefont
  {Ghiasi}}, \bibinfo {author} {\bibfnamefont {J.}~\bibnamefont
  {Ingla-Ayn{\'{e}}s}}, \bibinfo {author} {\bibfnamefont {A.~A.}\ \bibnamefont
  {Kaverzin}}, \ and\ \bibinfo {author} {\bibfnamefont {B.~J.}\ \bibnamefont
  {{van Wees}}},\ }\href {\doibase 10.1021/acs.nanolett.7b03460} {\bibfield
  {journal} {\bibinfo  {journal} {Nano Letters}\ }\textbf {\bibinfo {volume}
  {17}},\ \bibinfo {pages} {7528} (\bibinfo {year} {2017})}\BibitemShut
  {NoStop}%
\bibitem [{\citenamefont {Ben{\'{i}}tez}\ \emph {et~al.}(2017)\citenamefont
  {Ben{\'{i}}tez}, \citenamefont {Sierra}, \citenamefont {{Savero Torres}},
  \citenamefont {Arrighi}, \citenamefont {Bonell}, \citenamefont {Costache},\
  and\ \citenamefont {Valenzuela}}]{Benitez2017}%
  \BibitemOpen
  \bibfield  {author} {\bibinfo {author} {\bibfnamefont {L.~A.}\ \bibnamefont
  {Ben{\'{i}}tez}}, \bibinfo {author} {\bibfnamefont {J.~F.}\ \bibnamefont
  {Sierra}}, \bibinfo {author} {\bibfnamefont {W.}~\bibnamefont {{Savero
  Torres}}}, \bibinfo {author} {\bibfnamefont {A.}~\bibnamefont {Arrighi}},
  \bibinfo {author} {\bibfnamefont {F.}~\bibnamefont {Bonell}}, \bibinfo
  {author} {\bibfnamefont {M.~V.}\ \bibnamefont {Costache}}, \ and\ \bibinfo
  {author} {\bibfnamefont {S.~O.}\ \bibnamefont {Valenzuela}},\ }\href@noop {}
  {\bibfield  {journal} {\bibinfo  {journal} {Nature Physics}\ }\textbf
  {\bibinfo {volume} {14}},\ \bibinfo {pages} {1} (\bibinfo {year}
  {2017})}\BibitemShut {NoStop}%
\bibitem [{\citenamefont {Wang}\ \emph {et~al.}(2015)\citenamefont {Wang},
  \citenamefont {Ki}, \citenamefont {Chen}, \citenamefont {Berger},
  \citenamefont {MacDonald},\ and\ \citenamefont {Morpurgo}}]{Wang2015a}%
  \BibitemOpen
  \bibfield  {author} {\bibinfo {author} {\bibfnamefont {Z.}~\bibnamefont
  {Wang}}, \bibinfo {author} {\bibfnamefont {D.-K.}\ \bibnamefont {Ki}},
  \bibinfo {author} {\bibfnamefont {H.}~\bibnamefont {Chen}}, \bibinfo {author}
  {\bibfnamefont {H.}~\bibnamefont {Berger}}, \bibinfo {author} {\bibfnamefont
  {A.~H.}\ \bibnamefont {MacDonald}}, \ and\ \bibinfo {author} {\bibfnamefont
  {A.~F.}\ \bibnamefont {Morpurgo}},\ }\href
  {http://www.nature.com/ncomms/2015/150922/ncomms9339/full/ncomms9339.html}
  {\bibfield  {journal} {\bibinfo  {journal} {Nature communications}\ }\textbf
  {\bibinfo {volume} {6}},\ \bibinfo {pages} {8339} (\bibinfo {year}
  {2015})}\BibitemShut {NoStop}%
\bibitem [{\citenamefont {Wang}\ \emph {et~al.}(2016)\citenamefont {Wang},
  \citenamefont {Ki}, \citenamefont {Khoo}, \citenamefont {Mauro},
  \citenamefont {Berger}, \citenamefont {Levitov},\ and\ \citenamefont
  {Morpurgo}}]{Wang2016a}%
  \BibitemOpen
  \bibfield  {author} {\bibinfo {author} {\bibfnamefont {Z.}~\bibnamefont
  {Wang}}, \bibinfo {author} {\bibfnamefont {D.~K.}\ \bibnamefont {Ki}},
  \bibinfo {author} {\bibfnamefont {J.~Y.}\ \bibnamefont {Khoo}}, \bibinfo
  {author} {\bibfnamefont {D.}~\bibnamefont {Mauro}}, \bibinfo {author}
  {\bibfnamefont {H.}~\bibnamefont {Berger}}, \bibinfo {author} {\bibfnamefont
  {L.~S.}\ \bibnamefont {Levitov}}, \ and\ \bibinfo {author} {\bibfnamefont
  {A.~F.}\ \bibnamefont {Morpurgo}},\ }\href {\doibase
  10.1103/PhysRevX.6.041020} {\bibfield  {journal} {\bibinfo  {journal}
  {Physical Review X}\ }\textbf {\bibinfo {volume} {6}},\ \bibinfo {pages} {1}
  (\bibinfo {year} {2016})}\BibitemShut {NoStop}%
\bibitem [{\citenamefont {Zihlmann}\ \emph {et~al.}(2018)\citenamefont
  {Zihlmann}, \citenamefont {Cummings}, \citenamefont {Garcia}, \citenamefont
  {Kedves}, \citenamefont {Watanabe}, \citenamefont {Taniguchi}, \citenamefont
  {Sch{\"{o}}nenberger},\ and\ \citenamefont {Makk}}]{Zihlmann2018}%
  \BibitemOpen
  \bibfield  {author} {\bibinfo {author} {\bibfnamefont {S.}~\bibnamefont
  {Zihlmann}}, \bibinfo {author} {\bibfnamefont {A.~W.}\ \bibnamefont
  {Cummings}}, \bibinfo {author} {\bibfnamefont {J.~H.}\ \bibnamefont
  {Garcia}}, \bibinfo {author} {\bibfnamefont {M.}~\bibnamefont {Kedves}},
  \bibinfo {author} {\bibfnamefont {K.}~\bibnamefont {Watanabe}}, \bibinfo
  {author} {\bibfnamefont {T.}~\bibnamefont {Taniguchi}}, \bibinfo {author}
  {\bibfnamefont {C.}~\bibnamefont {Sch{\"{o}}nenberger}}, \ and\ \bibinfo
  {author} {\bibfnamefont {P.}~\bibnamefont {Makk}},\ }\href {\doibase
  10.1103/PhysRevB.97.075434} {\bibfield  {journal} {\bibinfo  {journal}
  {Physical Review B}\ }\textbf {\bibinfo {volume} {97}},\ \bibinfo {pages}
  {075434} (\bibinfo {year} {2018})}\BibitemShut {NoStop}%
\bibitem [{\citenamefont {Konschuh}\ \emph {et~al.}(2012)\citenamefont
  {Konschuh}, \citenamefont {Gmitra}, \citenamefont {Kochan},\ and\
  \citenamefont {Fabian}}]{Konschuh2012}%
  \BibitemOpen
  \bibfield  {author} {\bibinfo {author} {\bibfnamefont {S.}~\bibnamefont
  {Konschuh}}, \bibinfo {author} {\bibfnamefont {M.}~\bibnamefont {Gmitra}},
  \bibinfo {author} {\bibfnamefont {D.}~\bibnamefont {Kochan}}, \ and\ \bibinfo
  {author} {\bibfnamefont {J.}~\bibnamefont {Fabian}},\ }\href {\doibase
  10.1103/PhysRevB.85.115423} {\bibfield  {journal} {\bibinfo  {journal}
  {Physical Review B - Condensed Matter and Materials Physics}\ }\textbf
  {\bibinfo {volume} {85}},\ \bibinfo {pages} {1} (\bibinfo {year}
  {2012})}\BibitemShut {NoStop}%
\bibitem [{\citenamefont {Gmitra}\ \emph {et~al.}(2018)\citenamefont {Gmitra},
  \citenamefont {Zollner},\ and\ \citenamefont {Fabian}}]{Fabiannew}%
  \BibitemOpen
  \bibfield  {author} {\bibinfo {author} {\bibfnamefont {M.}~\bibnamefont
  {Gmitra}}, \bibinfo {author} {\bibfnamefont {K.}~\bibnamefont {Zollner}}, \
  and\ \bibinfo {author} {\bibfnamefont {J.}~\bibnamefont {Fabian}},\
  }\href@noop {} {\bibfield  {journal} {\bibinfo  {journal} {in preparation}\ }
  (\bibinfo {year} {2018})}\BibitemShut {NoStop}%
\bibitem [{\citenamefont {Martin}\ \emph {et~al.}(2008)\citenamefont {Martin},
  \citenamefont {Akerman}, \citenamefont {Ulbricht}, \citenamefont {Lohmann},
  \citenamefont {Smet}, \citenamefont {{Von Klitzing}},\ and\ \citenamefont
  {Yacoby}}]{Martin2008}%
  \BibitemOpen
  \bibfield  {author} {\bibinfo {author} {\bibfnamefont {J.}~\bibnamefont
  {Martin}}, \bibinfo {author} {\bibfnamefont {N.}~\bibnamefont {Akerman}},
  \bibinfo {author} {\bibfnamefont {G.}~\bibnamefont {Ulbricht}}, \bibinfo
  {author} {\bibfnamefont {T.}~\bibnamefont {Lohmann}}, \bibinfo {author}
  {\bibfnamefont {J.~H.}\ \bibnamefont {Smet}}, \bibinfo {author}
  {\bibfnamefont {K.}~\bibnamefont {{Von Klitzing}}}, \ and\ \bibinfo {author}
  {\bibfnamefont {A.}~\bibnamefont {Yacoby}},\ }\href {\doibase
  10.1038/nphys781} {\bibfield  {journal} {\bibinfo  {journal} {Nature
  Physics}\ }\textbf {\bibinfo {volume} {4}},\ \bibinfo {pages} {144} (\bibinfo
  {year} {2008})}\BibitemShut {NoStop}%
\bibitem [{\citenamefont {Leutenantsmeyer}\ \emph {et~al.}(2017)\citenamefont
  {Leutenantsmeyer}, \citenamefont {Kaverzin}, \citenamefont {Wojtaszek},\ and\
  \citenamefont {van Wees}}]{Leutenantsmeyer2017}%
  \BibitemOpen
  \bibfield  {author} {\bibinfo {author} {\bibfnamefont {J.~C.}\ \bibnamefont
  {Leutenantsmeyer}}, \bibinfo {author} {\bibfnamefont {A.~A.}\ \bibnamefont
  {Kaverzin}}, \bibinfo {author} {\bibfnamefont {M.}~\bibnamefont {Wojtaszek}},
  \ and\ \bibinfo {author} {\bibfnamefont {B.~J.}\ \bibnamefont {van Wees}},\
  }\href {\doibase 10.1088/2053-1583/4/1/014001} {\bibfield  {journal}
  {\bibinfo  {journal} {2D Materials}\ }\textbf {\bibinfo {volume} {4}},\
  \bibinfo {pages} {014001} (\bibinfo {year} {2017})}\BibitemShut {NoStop}%
\bibitem [{\citenamefont {{Van Tuan}}\ \emph {et~al.}(2016)\citenamefont {{Van
  Tuan}}, \citenamefont {Adam},\ and\ \citenamefont {Roche}}]{VanTuan2016}%
  \BibitemOpen
  \bibfield  {author} {\bibinfo {author} {\bibfnamefont {D.}~\bibnamefont {{Van
  Tuan}}}, \bibinfo {author} {\bibfnamefont {S.}~\bibnamefont {Adam}}, \ and\
  \bibinfo {author} {\bibfnamefont {S.}~\bibnamefont {Roche}},\ }\href
  {\doibase 10.1103/PhysRevB.94.041405} {\bibfield  {journal} {\bibinfo
  {journal} {Physical Review B}\ }\textbf {\bibinfo {volume} {94}},\ \bibinfo
  {pages} {1} (\bibinfo {year} {2016})}\BibitemShut {NoStop}%
\bibitem [{\citenamefont {Rashba}(2009)}]{Rashba2009}%
  \BibitemOpen
  \bibfield  {author} {\bibinfo {author} {\bibfnamefont {E.~I.}\ \bibnamefont
  {Rashba}},\ }\href {\doibase 10.1103/PhysRevB.79.161409} {\bibfield
  {journal} {\bibinfo  {journal} {Physical Review B - Condensed Matter and
  Materials Physics}\ }\textbf {\bibinfo {volume} {79}},\ \bibinfo {pages} {1}
  (\bibinfo {year} {2009})}\BibitemShut {NoStop}%
\bibitem [{\citenamefont {Guimar{\~{a}}es}\ \emph {et~al.}(2014)\citenamefont
  {Guimar{\~{a}}es}, \citenamefont {Zomer}, \citenamefont {Ingla-Ayn{\'{e}}s},
  \citenamefont {Brant}, \citenamefont {Tombros},\ and\ \citenamefont {van
  Wees}}]{Guimaraes2014}%
  \BibitemOpen
  \bibfield  {author} {\bibinfo {author} {\bibfnamefont {M.~H.~D.}\
  \bibnamefont {Guimar{\~{a}}es}}, \bibinfo {author} {\bibfnamefont {P.~J.}\
  \bibnamefont {Zomer}}, \bibinfo {author} {\bibfnamefont {J.}~\bibnamefont
  {Ingla-Ayn{\'{e}}s}}, \bibinfo {author} {\bibfnamefont {J.~C.}\ \bibnamefont
  {Brant}}, \bibinfo {author} {\bibfnamefont {N.}~\bibnamefont {Tombros}}, \
  and\ \bibinfo {author} {\bibfnamefont {B.~J.}\ \bibnamefont {van Wees}},\
  }\href@noop {} {\bibfield  {journal} {\bibinfo  {journal} {Physical Review
  Letters}\ }\textbf {\bibinfo {volume} {113}},\ \bibinfo {pages} {1} (\bibinfo
  {year} {2014})}\BibitemShut {NoStop}%
\bibitem [{\citenamefont {Wang}\ and\ \citenamefont {Wu}(2013)}]{Wang2013}%
  \BibitemOpen
  \bibfield  {author} {\bibinfo {author} {\bibfnamefont {L.}~\bibnamefont
  {Wang}}\ and\ \bibinfo {author} {\bibfnamefont {M.~W.}\ \bibnamefont {Wu}},\
  }\href {\doibase 10.1103/PhysRevB.87.205416} {\bibfield  {journal} {\bibinfo
  {journal} {Physical Review B}\ }\textbf {\bibinfo {volume} {87}},\ \bibinfo
  {pages} {205416} (\bibinfo {year} {2013})}\BibitemShut {NoStop}%
\bibitem [{\citenamefont {Tombros}\ \emph {et~al.}(2008)\citenamefont
  {Tombros}, \citenamefont {Tanabe}, \citenamefont {Veligura}, \citenamefont
  {Jozsa}, \citenamefont {Popinciuc}, \citenamefont {Jonkman},\ and\
  \citenamefont {{van Wees}}}]{Tombros2008}%
  \BibitemOpen
  \bibfield  {author} {\bibinfo {author} {\bibfnamefont {N.}~\bibnamefont
  {Tombros}}, \bibinfo {author} {\bibfnamefont {S.}~\bibnamefont {Tanabe}},
  \bibinfo {author} {\bibfnamefont {A.}~\bibnamefont {Veligura}}, \bibinfo
  {author} {\bibfnamefont {C.}~\bibnamefont {Jozsa}}, \bibinfo {author}
  {\bibfnamefont {M.}~\bibnamefont {Popinciuc}}, \bibinfo {author}
  {\bibfnamefont {H.~T.}\ \bibnamefont {Jonkman}}, \ and\ \bibinfo {author}
  {\bibfnamefont {B.~J.}\ \bibnamefont {{van Wees}}},\ }\href {\doibase
  10.1103/PhysRevLett.101.046601} {\bibfield  {journal} {\bibinfo  {journal}
  {Physical Review Letters}\ }\textbf {\bibinfo {volume} {101}},\ \bibinfo
  {pages} {2} (\bibinfo {year} {2008})}\BibitemShut {NoStop}%
\bibitem [{\citenamefont {Raes}\ \emph {et~al.}(2016)\citenamefont {Raes},
  \citenamefont {Scheerder}, \citenamefont {Costache}, \citenamefont {Bonell},
  \citenamefont {Sierra}, \citenamefont {Cuppens}, \citenamefont {{van de
  Vondel}},\ and\ \citenamefont {Valenzuela}}]{Raes2016}%
  \BibitemOpen
  \bibfield  {author} {\bibinfo {author} {\bibfnamefont {B.}~\bibnamefont
  {Raes}}, \bibinfo {author} {\bibfnamefont {J.~E.}\ \bibnamefont {Scheerder}},
  \bibinfo {author} {\bibfnamefont {M.~V.}\ \bibnamefont {Costache}}, \bibinfo
  {author} {\bibfnamefont {F.}~\bibnamefont {Bonell}}, \bibinfo {author}
  {\bibfnamefont {J.~F.}\ \bibnamefont {Sierra}}, \bibinfo {author}
  {\bibfnamefont {J.}~\bibnamefont {Cuppens}}, \bibinfo {author} {\bibfnamefont
  {J.}~\bibnamefont {{van de Vondel}}}, \ and\ \bibinfo {author} {\bibfnamefont
  {S.~O.}\ \bibnamefont {Valenzuela}},\ }\href {\doibase 10.1038/ncomms11444}
  {\bibfield  {journal} {\bibinfo  {journal} {Nature Communications}\ }\textbf
  {\bibinfo {volume} {7}},\ \bibinfo {pages} {11444} (\bibinfo {year}
  {2016})}\BibitemShut {NoStop}%
\bibitem [{\citenamefont {Ringer}\ \emph {et~al.}(2018)\citenamefont {Ringer},
  \citenamefont {Hartl}, \citenamefont {Rosenauer}, \citenamefont
  {V{\"{o}}lkl}, \citenamefont {Kadur}, \citenamefont {Hopperdietzel},
  \citenamefont {Weiss},\ and\ \citenamefont {Eroms}}]{Ringer2017}%
  \BibitemOpen
  \bibfield  {author} {\bibinfo {author} {\bibfnamefont {S.}~\bibnamefont
  {Ringer}}, \bibinfo {author} {\bibfnamefont {S.}~\bibnamefont {Hartl}},
  \bibinfo {author} {\bibfnamefont {M.}~\bibnamefont {Rosenauer}}, \bibinfo
  {author} {\bibfnamefont {T.}~\bibnamefont {V{\"{o}}lkl}}, \bibinfo {author}
  {\bibfnamefont {M.}~\bibnamefont {Kadur}}, \bibinfo {author} {\bibfnamefont
  {F.}~\bibnamefont {Hopperdietzel}}, \bibinfo {author} {\bibfnamefont
  {D.}~\bibnamefont {Weiss}}, \ and\ \bibinfo {author} {\bibfnamefont
  {J.}~\bibnamefont {Eroms}},\ }\href {\doibase 10.1103/PhysRevB.97.205439}
  {\bibfield  {journal} {\bibinfo  {journal} {Physical Review B}\ }\textbf
  {\bibinfo {volume} {97}},\ \bibinfo {pages} {205439} (\bibinfo {year}
  {2018})}\BibitemShut {NoStop}%
\bibitem [{\citenamefont {Han}\ and\ \citenamefont {Kawakami}(2011)}]{Han2011}%
  \BibitemOpen
  \bibfield  {author} {\bibinfo {author} {\bibfnamefont {W.}~\bibnamefont
  {Han}}\ and\ \bibinfo {author} {\bibfnamefont {R.~K.}\ \bibnamefont
  {Kawakami}},\ }\href {\doibase 10.1103/PhysRevLett.107.047207} {\bibfield
  {journal} {\bibinfo  {journal} {Physical Review Letters}\ }\textbf {\bibinfo
  {volume} {107}},\ \bibinfo {pages} {1} (\bibinfo {year} {2011})}\BibitemShut
  {NoStop}%
\bibitem [{\citenamefont {Yang}\ \emph {et~al.}(2011)\citenamefont {Yang},
  \citenamefont {Balakrishnan}, \citenamefont {Volmer}, \citenamefont {Avsar},
  \citenamefont {Jaiswal}, \citenamefont {Samm}, \citenamefont {Ali},
  \citenamefont {Pachoud}, \citenamefont {Zeng}, \citenamefont {Popinciuc},
  \citenamefont {G{\"{u}}ntherodt}, \citenamefont {Beschoten},\ and\
  \citenamefont {{\"{O}}zyilmaz}}]{Yang2011}%
  \BibitemOpen
  \bibfield  {author} {\bibinfo {author} {\bibfnamefont {T.~Y.}\ \bibnamefont
  {Yang}}, \bibinfo {author} {\bibfnamefont {J.}~\bibnamefont {Balakrishnan}},
  \bibinfo {author} {\bibfnamefont {F.}~\bibnamefont {Volmer}}, \bibinfo
  {author} {\bibfnamefont {A.}~\bibnamefont {Avsar}}, \bibinfo {author}
  {\bibfnamefont {M.}~\bibnamefont {Jaiswal}}, \bibinfo {author} {\bibfnamefont
  {J.}~\bibnamefont {Samm}}, \bibinfo {author} {\bibfnamefont {S.~R.}\
  \bibnamefont {Ali}}, \bibinfo {author} {\bibfnamefont {A.}~\bibnamefont
  {Pachoud}}, \bibinfo {author} {\bibfnamefont {M.}~\bibnamefont {Zeng}},
  \bibinfo {author} {\bibfnamefont {M.}~\bibnamefont {Popinciuc}}, \bibinfo
  {author} {\bibfnamefont {G.}~\bibnamefont {G{\"{u}}ntherodt}}, \bibinfo
  {author} {\bibfnamefont {B.}~\bibnamefont {Beschoten}}, \ and\ \bibinfo
  {author} {\bibfnamefont {B.}~\bibnamefont {{\"{O}}zyilmaz}},\ }\href
  {\doibase 10.1103/PhysRevLett.107.047206} {\bibfield  {journal} {\bibinfo
  {journal} {Physical Review Letters}\ }\textbf {\bibinfo {volume} {107}},\
  \bibinfo {pages} {5} (\bibinfo {year} {2011})}\BibitemShut {NoStop}%
\bibitem [{\citenamefont {Avsar}\ \emph {et~al.}(2011)\citenamefont {Avsar},
  \citenamefont {Yang}, \citenamefont {Bae}, \citenamefont {Balakrishnan},
  \citenamefont {Volmer}, \citenamefont {Jaiswal}, \citenamefont {Yi},
  \citenamefont {Ali}, \citenamefont {G{\"{u}}ntherodt}, \citenamefont {Hong},
  \citenamefont {Beschoten},\ and\ \citenamefont {{\"{O}}zyilmaz}}]{Avsar2011}%
  \BibitemOpen
  \bibfield  {author} {\bibinfo {author} {\bibfnamefont {A.}~\bibnamefont
  {Avsar}}, \bibinfo {author} {\bibfnamefont {T.~Y.}\ \bibnamefont {Yang}},
  \bibinfo {author} {\bibfnamefont {S.}~\bibnamefont {Bae}}, \bibinfo {author}
  {\bibfnamefont {J.}~\bibnamefont {Balakrishnan}}, \bibinfo {author}
  {\bibfnamefont {F.}~\bibnamefont {Volmer}}, \bibinfo {author} {\bibfnamefont
  {M.}~\bibnamefont {Jaiswal}}, \bibinfo {author} {\bibfnamefont
  {Z.}~\bibnamefont {Yi}}, \bibinfo {author} {\bibfnamefont {S.~R.}\
  \bibnamefont {Ali}}, \bibinfo {author} {\bibfnamefont {G.}~\bibnamefont
  {G{\"{u}}ntherodt}}, \bibinfo {author} {\bibfnamefont {B.~H.}\ \bibnamefont
  {Hong}}, \bibinfo {author} {\bibfnamefont {B.}~\bibnamefont {Beschoten}}, \
  and\ \bibinfo {author} {\bibfnamefont {B.}~\bibnamefont {{\"{O}}zyilmaz}},\
  }\href {\doibase 10.1021/nl200714q} {\bibfield  {journal} {\bibinfo
  {journal} {Nano Letters}\ }\textbf {\bibinfo {volume} {11}},\ \bibinfo
  {pages} {2363} (\bibinfo {year} {2011})}\BibitemShut {NoStop}%
\bibitem [{\citenamefont {Neumann}\ \emph {et~al.}(2013)\citenamefont
  {Neumann}, \citenamefont {{Van De Vondel}}, \citenamefont {Bridoux},
  \citenamefont {Costache}, \citenamefont {Alzina}, \citenamefont {{Sotomayor
  Torres}},\ and\ \citenamefont {Valenzuela}}]{Neumann2013a}%
  \BibitemOpen
  \bibfield  {author} {\bibinfo {author} {\bibfnamefont {I.}~\bibnamefont
  {Neumann}}, \bibinfo {author} {\bibfnamefont {J.}~\bibnamefont {{Van De
  Vondel}}}, \bibinfo {author} {\bibfnamefont {G.}~\bibnamefont {Bridoux}},
  \bibinfo {author} {\bibfnamefont {M.~V.}\ \bibnamefont {Costache}}, \bibinfo
  {author} {\bibfnamefont {F.}~\bibnamefont {Alzina}}, \bibinfo {author}
  {\bibfnamefont {C.~M.}\ \bibnamefont {{Sotomayor Torres}}}, \ and\ \bibinfo
  {author} {\bibfnamefont {S.~O.}\ \bibnamefont {Valenzuela}},\ }\href
  {\doibase 10.1002/smll.201201194} {\bibfield  {journal} {\bibinfo  {journal}
  {Small}\ }\textbf {\bibinfo {volume} {9}},\ \bibinfo {pages} {156} (\bibinfo
  {year} {2013})}\BibitemShut {NoStop}%
\bibitem [{\citenamefont {Ingla-Ayn{\'{e}}s}\ \emph {et~al.}(2015)\citenamefont
  {Ingla-Ayn{\'{e}}s}, \citenamefont {Guimar{\~{a}}es}, \citenamefont
  {Meijerink}, \citenamefont {Zomer},\ and\ \citenamefont {{van
  Wees}}}]{Ingla-Aynes2015}%
  \BibitemOpen
  \bibfield  {author} {\bibinfo {author} {\bibfnamefont {J.}~\bibnamefont
  {Ingla-Ayn{\'{e}}s}}, \bibinfo {author} {\bibfnamefont {M.~H.}\ \bibnamefont
  {Guimar{\~{a}}es}}, \bibinfo {author} {\bibfnamefont {R.~J.}\ \bibnamefont
  {Meijerink}}, \bibinfo {author} {\bibfnamefont {P.~J.}\ \bibnamefont
  {Zomer}}, \ and\ \bibinfo {author} {\bibfnamefont {B.~J.}\ \bibnamefont {{van
  Wees}}},\ }\href {\doibase 10.1103/PhysRevB.92.201410} {\bibfield  {journal}
  {\bibinfo  {journal} {Physical Review B}\ }\textbf {\bibinfo {volume} {92}},\
  \bibinfo {pages} {1} (\bibinfo {year} {2015})}\BibitemShut {NoStop}%
\bibitem [{\citenamefont {Avsar}\ \emph {et~al.}(2016)\citenamefont {Avsar},
  \citenamefont {Vera-Marun}, \citenamefont {Tan}, \citenamefont {Koon},
  \citenamefont {Watanabe}, \citenamefont {Taniguchi}, \citenamefont {Adam},\
  and\ \citenamefont {{\"{O}}zyilmaz}}]{Avsar2016}%
  \BibitemOpen
  \bibfield  {author} {\bibinfo {author} {\bibfnamefont {A.}~\bibnamefont
  {Avsar}}, \bibinfo {author} {\bibfnamefont {I.~J.}\ \bibnamefont
  {Vera-Marun}}, \bibinfo {author} {\bibfnamefont {J.~Y.}\ \bibnamefont {Tan}},
  \bibinfo {author} {\bibfnamefont {G.~K.~W.}\ \bibnamefont {Koon}}, \bibinfo
  {author} {\bibfnamefont {K.}~\bibnamefont {Watanabe}}, \bibinfo {author}
  {\bibfnamefont {T.}~\bibnamefont {Taniguchi}}, \bibinfo {author}
  {\bibfnamefont {S.}~\bibnamefont {Adam}}, \ and\ \bibinfo {author}
  {\bibfnamefont {B.}~\bibnamefont {{\"{O}}zyilmaz}},\ }\href@noop {}
  {\bibfield  {journal} {\bibinfo  {journal} {NPG Asia Materials}\ }\textbf
  {\bibinfo {volume} {8}} (\bibinfo {year} {2016})}\BibitemShut {NoStop}%
\bibitem [{\citenamefont {Zomer}\ \emph {et~al.}(2014)\citenamefont {Zomer},
  \citenamefont {Guimar{\~{a}}es}, \citenamefont {Brant}, \citenamefont
  {Tombros},\ and\ \citenamefont {van Wees}}]{Zomer2014}%
  \BibitemOpen
  \bibfield  {author} {\bibinfo {author} {\bibfnamefont {P.~J.}\ \bibnamefont
  {Zomer}}, \bibinfo {author} {\bibfnamefont {M.~H.~D.}\ \bibnamefont
  {Guimar{\~{a}}es}}, \bibinfo {author} {\bibfnamefont {J.~C.}\ \bibnamefont
  {Brant}}, \bibinfo {author} {\bibfnamefont {N.}~\bibnamefont {Tombros}}, \
  and\ \bibinfo {author} {\bibfnamefont {B.~J.}\ \bibnamefont {van Wees}},\
  }\href {\doibase 10.1063/1.4886096} {\bibfield  {journal} {\bibinfo
  {journal} {Applied Physics Letters}\ }\textbf {\bibinfo {volume} {105}},\
  \bibinfo {pages} {013101} (\bibinfo {year} {2014})}\BibitemShut {NoStop}%
\bibitem [{Note1()}]{Note1}%
  \BibitemOpen
  \bibinfo {note} {See Supplemental Material.}\BibitemShut {Stop}%
\bibitem [{\citenamefont {Castro}\ \emph {et~al.}(2007)\citenamefont {Castro},
  \citenamefont {Novoselov}, \citenamefont {Morozov}, \citenamefont {Peres},
  \citenamefont {{Lopes dos Santos}}, \citenamefont {Nilsson}, \citenamefont
  {Guinea}, \citenamefont {Geim},\ and\ \citenamefont {{Castro
  Neto}}}]{Castro2007}%
  \BibitemOpen
  \bibfield  {author} {\bibinfo {author} {\bibfnamefont {E.~V.}\ \bibnamefont
  {Castro}}, \bibinfo {author} {\bibfnamefont {K.~S.}\ \bibnamefont
  {Novoselov}}, \bibinfo {author} {\bibfnamefont {S.~V.}\ \bibnamefont
  {Morozov}}, \bibinfo {author} {\bibfnamefont {N.~M.~R.}\ \bibnamefont
  {Peres}}, \bibinfo {author} {\bibfnamefont {J.~M. B.~.}\ \bibnamefont {{Lopes
  dos Santos}}}, \bibinfo {author} {\bibfnamefont {J.}~\bibnamefont {Nilsson}},
  \bibinfo {author} {\bibfnamefont {F.}~\bibnamefont {Guinea}}, \bibinfo
  {author} {\bibfnamefont {A.~K.}\ \bibnamefont {Geim}}, \ and\ \bibinfo
  {author} {\bibfnamefont {A.~H.}\ \bibnamefont {{Castro Neto}}},\ }\href
  {\doibase 10.1103/PhysRevLett.99.216802} {\bibfield  {journal} {\bibinfo
  {journal} {Physical Review Letters}\ }\textbf {\bibinfo {volume} {99}},\
  \bibinfo {pages} {8} (\bibinfo {year} {2007})}\BibitemShut {NoStop}%
\bibitem [{\citenamefont {Oostinga}\ \emph {et~al.}(2008)\citenamefont
  {Oostinga}, \citenamefont {Heersche}, \citenamefont {Liu}, \citenamefont
  {Morpurgo},\ and\ \citenamefont {Vandersypen}}]{Oostinga2008}%
  \BibitemOpen
  \bibfield  {author} {\bibinfo {author} {\bibfnamefont {J.~B.}\ \bibnamefont
  {Oostinga}}, \bibinfo {author} {\bibfnamefont {H.~B.}\ \bibnamefont
  {Heersche}}, \bibinfo {author} {\bibfnamefont {X.}~\bibnamefont {Liu}},
  \bibinfo {author} {\bibfnamefont {A.~F.}\ \bibnamefont {Morpurgo}}, \ and\
  \bibinfo {author} {\bibfnamefont {L.~M.~K.}\ \bibnamefont {Vandersypen}},\
  }\href {\doibase 10.1038/nmat2082} {\bibfield  {journal} {\bibinfo  {journal}
  {Nature Materials}\ }\textbf {\bibinfo {volume} {7}},\ \bibinfo {pages} {151}
  (\bibinfo {year} {2008})}\BibitemShut {NoStop}%
\bibitem [{\citenamefont {Zhang}\ \emph {et~al.}(2009)\citenamefont {Zhang},
  \citenamefont {Tang}, \citenamefont {Girit}, \citenamefont {Hao},
  \citenamefont {Martin}, \citenamefont {Zettl}, \citenamefont {Crommie},
  \citenamefont {{Ron Shen}},\ and\ \citenamefont {Wang}}]{Zhang2009}%
  \BibitemOpen
  \bibfield  {author} {\bibinfo {author} {\bibfnamefont {Y.}~\bibnamefont
  {Zhang}}, \bibinfo {author} {\bibfnamefont {T.-T.}\ \bibnamefont {Tang}},
  \bibinfo {author} {\bibfnamefont {C.}~\bibnamefont {Girit}}, \bibinfo
  {author} {\bibfnamefont {Z.}~\bibnamefont {Hao}}, \bibinfo {author}
  {\bibfnamefont {M.~C.}\ \bibnamefont {Martin}}, \bibinfo {author}
  {\bibfnamefont {A.}~\bibnamefont {Zettl}}, \bibinfo {author} {\bibfnamefont
  {M.~F.}\ \bibnamefont {Crommie}}, \bibinfo {author} {\bibfnamefont
  {Y.}~\bibnamefont {{Ron Shen}}}, \ and\ \bibinfo {author} {\bibfnamefont
  {F.}~\bibnamefont {Wang}},\ }\href {\doibase 10.1038/nature08105} {\bibfield
  {journal} {\bibinfo  {journal} {Nature}\ }\textbf {\bibinfo {volume} {459}},\
  \bibinfo {pages} {820} (\bibinfo {year} {2009})}\BibitemShut {NoStop}%
\bibitem [{\citenamefont {Gurram}\ \emph {et~al.}(2017)\citenamefont {Gurram},
  \citenamefont {Omar},\ and\ \citenamefont {van Wees}}]{Gurram2017}%
  \BibitemOpen
  \bibfield  {author} {\bibinfo {author} {\bibfnamefont {M.}~\bibnamefont
  {Gurram}}, \bibinfo {author} {\bibfnamefont {S.}~\bibnamefont {Omar}}, \ and\
  \bibinfo {author} {\bibfnamefont {B.~J.}\ \bibnamefont {van Wees}},\ }\href
  {\doibase 10.1038/s41467-017-00317-w} {\bibfield  {journal} {\bibinfo
  {journal} {Nature Communications}\ }\textbf {\bibinfo {volume} {8}},\
  \bibinfo {pages} {248} (\bibinfo {year} {2017})}\BibitemShut {NoStop}%
\bibitem [{\citenamefont {Gurram}\ \emph {et~al.}(2018)\citenamefont {Gurram},
  \citenamefont {Omar},\ and\ \citenamefont {{van Wees}}}]{Gurram2018}%
  \BibitemOpen
  \bibfield  {author} {\bibinfo {author} {\bibfnamefont {M.}~\bibnamefont
  {Gurram}}, \bibinfo {author} {\bibfnamefont {S.}~\bibnamefont {Omar}}, \ and\
  \bibinfo {author} {\bibfnamefont {B.~J.}\ \bibnamefont {{van Wees}}},\ }\href
  {http://arxiv.org/abs/1712.07828} {\bibfield  {journal} {\bibinfo  {journal}
  {2D Materials}\ } (\bibinfo {year} {2018})},\ \Eprint
  {http://arxiv.org/abs/1712.07828} {arXiv:1712.07828} \BibitemShut {NoStop}%
\bibitem [{\citenamefont {Raes}\ \emph {et~al.}(2017)\citenamefont {Raes},
  \citenamefont {Cummings}, \citenamefont {Bonell}, \citenamefont {Costache},
  \citenamefont {Sierra}, \citenamefont {Roche},\ and\ \citenamefont
  {Valenzuela}}]{Raes2017}%
  \BibitemOpen
  \bibfield  {author} {\bibinfo {author} {\bibfnamefont {B.}~\bibnamefont
  {Raes}}, \bibinfo {author} {\bibfnamefont {A.~W.}\ \bibnamefont {Cummings}},
  \bibinfo {author} {\bibfnamefont {F.}~\bibnamefont {Bonell}}, \bibinfo
  {author} {\bibfnamefont {M.~V.}\ \bibnamefont {Costache}}, \bibinfo {author}
  {\bibfnamefont {J.~F.}\ \bibnamefont {Sierra}}, \bibinfo {author}
  {\bibfnamefont {S.}~\bibnamefont {Roche}}, \ and\ \bibinfo {author}
  {\bibfnamefont {S.~O.}\ \bibnamefont {Valenzuela}},\ }\href {\doibase
  10.1103/PhysRevB.95.085403} {\bibfield  {journal} {\bibinfo  {journal}
  {Physical Review B}\ }\textbf {\bibinfo {volume} {95}},\ \bibinfo {pages} {1}
  (\bibinfo {year} {2017})}\BibitemShut {NoStop}%
\bibitem [{\citenamefont {Maassen}\ \emph {et~al.}(2012)\citenamefont
  {Maassen}, \citenamefont {Vera-Marun}, \citenamefont {Guimar{\~{a}}es},\ and\
  \citenamefont {van Wees}}]{Maassen2012}%
  \BibitemOpen
  \bibfield  {author} {\bibinfo {author} {\bibfnamefont {T.}~\bibnamefont
  {Maassen}}, \bibinfo {author} {\bibfnamefont {I.~J.}\ \bibnamefont
  {Vera-Marun}}, \bibinfo {author} {\bibfnamefont {M.~H.~D.}\ \bibnamefont
  {Guimar{\~{a}}es}}, \ and\ \bibinfo {author} {\bibfnamefont {B.~J.}\
  \bibnamefont {van Wees}},\ }\href {\doibase 10.1103/PhysRevB.86.235408}
  {\bibfield  {journal} {\bibinfo  {journal} {Physical Review B}\ }\textbf
  {\bibinfo {volume} {86}},\ \bibinfo {pages} {235408} (\bibinfo {year}
  {2012})}\BibitemShut {NoStop}%
\bibitem [{\citenamefont {Zhu}\ and\ \citenamefont {Kawakami}(2018)}]{Zhu2018}%
  \BibitemOpen
  \bibfield  {author} {\bibinfo {author} {\bibfnamefont {T.}~\bibnamefont
  {Zhu}}\ and\ \bibinfo {author} {\bibfnamefont {R.~K.}\ \bibnamefont
  {Kawakami}},\ }\href {\doibase 10.1103/PhysRevB.97.144413} {\bibfield
  {journal} {\bibinfo  {journal} {Physical Review B}\ }\textbf {\bibinfo
  {volume} {97}},\ \bibinfo {pages} {144413} (\bibinfo {year}
  {2018})}\BibitemShut {NoStop}%
\bibitem [{\citenamefont {Dr{\"{o}}geler}\ \emph {et~al.}(2016)\citenamefont
  {Dr{\"{o}}geler}, \citenamefont {Franzen}, \citenamefont {Volmer},
  \citenamefont {Pohlmann}, \citenamefont {Banszerus}, \citenamefont {Wolter},
  \citenamefont {Watanabe}, \citenamefont {Taniguchi}, \citenamefont
  {Stampfer},\ and\ \citenamefont {Beschoten}}]{Drogeler2016}%
  \BibitemOpen
  \bibfield  {author} {\bibinfo {author} {\bibfnamefont {M.}~\bibnamefont
  {Dr{\"{o}}geler}}, \bibinfo {author} {\bibfnamefont {C.}~\bibnamefont
  {Franzen}}, \bibinfo {author} {\bibfnamefont {F.}~\bibnamefont {Volmer}},
  \bibinfo {author} {\bibfnamefont {T.}~\bibnamefont {Pohlmann}}, \bibinfo
  {author} {\bibfnamefont {L.}~\bibnamefont {Banszerus}}, \bibinfo {author}
  {\bibfnamefont {M.}~\bibnamefont {Wolter}}, \bibinfo {author} {\bibfnamefont
  {K.}~\bibnamefont {Watanabe}}, \bibinfo {author} {\bibfnamefont
  {T.}~\bibnamefont {Taniguchi}}, \bibinfo {author} {\bibfnamefont
  {C.}~\bibnamefont {Stampfer}}, \ and\ \bibinfo {author} {\bibfnamefont
  {B.}~\bibnamefont {Beschoten}},\ }\href {\doibase
  10.1021/acs.nanolett.6b00497} {\bibfield  {journal} {\bibinfo  {journal}
  {Nano Letters}\ }\textbf {\bibinfo {volume} {16}},\ \bibinfo {pages} {3533}
  (\bibinfo {year} {2016})}\BibitemShut {NoStop}%
\bibitem [{\citenamefont {Kochan}\ \emph {et~al.}(2015)\citenamefont {Kochan},
  \citenamefont {Irmer}, \citenamefont {Gmitra},\ and\ \citenamefont
  {Fabian}}]{Kochan2015}%
  \BibitemOpen
  \bibfield  {author} {\bibinfo {author} {\bibfnamefont {D.}~\bibnamefont
  {Kochan}}, \bibinfo {author} {\bibfnamefont {S.}~\bibnamefont {Irmer}},
  \bibinfo {author} {\bibfnamefont {M.}~\bibnamefont {Gmitra}}, \ and\ \bibinfo
  {author} {\bibfnamefont {J.}~\bibnamefont {Fabian}},\ }\href {\doibase
  10.1103/PhysRevLett.115.196601} {\bibfield  {journal} {\bibinfo  {journal}
  {Physical Review Letters}\ }\textbf {\bibinfo {volume} {115}},\ \bibinfo
  {pages} {6} (\bibinfo {year} {2015})}\BibitemShut {NoStop}%
\bibitem [{\citenamefont {{Van Tuan}}\ \emph {et~al.}(2014)\citenamefont {{Van
  Tuan}}, \citenamefont {Ortmann}, \citenamefont {Soriano}, \citenamefont
  {Valenzuela},\ and\ \citenamefont {Roche}}]{Tuan2014}%
  \BibitemOpen
  \bibfield  {author} {\bibinfo {author} {\bibfnamefont {D.}~\bibnamefont {{Van
  Tuan}}}, \bibinfo {author} {\bibfnamefont {F.}~\bibnamefont {Ortmann}},
  \bibinfo {author} {\bibfnamefont {D.}~\bibnamefont {Soriano}}, \bibinfo
  {author} {\bibfnamefont {S.~O.}\ \bibnamefont {Valenzuela}}, \ and\ \bibinfo
  {author} {\bibfnamefont {S.}~\bibnamefont {Roche}},\ }\href@noop {}
  {\bibfield  {journal} {\bibinfo  {journal} {Nature Physics}\ }\textbf
  {\bibinfo {volume} {10}},\ \bibinfo {pages} {857} (\bibinfo {year}
  {2014})}\BibitemShut {NoStop}%
\bibitem [{\citenamefont {Han}\ \emph {et~al.}(2014)\citenamefont {Han},
  \citenamefont {Kawakami}, \citenamefont {Gmitra},\ and\ \citenamefont
  {Fabian}}]{Han2014}%
  \BibitemOpen
  \bibfield  {author} {\bibinfo {author} {\bibfnamefont {W.}~\bibnamefont
  {Han}}, \bibinfo {author} {\bibfnamefont {R.~K.}\ \bibnamefont {Kawakami}},
  \bibinfo {author} {\bibfnamefont {M.}~\bibnamefont {Gmitra}}, \ and\ \bibinfo
  {author} {\bibfnamefont {J.}~\bibnamefont {Fabian}},\ }\href {\doibase
  10.1038/nnano.2014.214} {\bibfield  {journal} {\bibinfo  {journal} {Nature
  Nanotechnology}\ }\textbf {\bibinfo {volume} {9}},\ \bibinfo {pages} {794}
  (\bibinfo {year} {2014})}\BibitemShut {NoStop}%
\bibitem [{\citenamefont {Fabian}\ \emph {et~al.}(2007)\citenamefont {Fabian},
  \citenamefont {Matos-Abiague}, \citenamefont {Ertler}, \citenamefont
  {Stano},\ and\ \citenamefont {Zutic}}]{Fabian2007}%
  \BibitemOpen
  \bibfield  {author} {\bibinfo {author} {\bibfnamefont {J.}~\bibnamefont
  {Fabian}}, \bibinfo {author} {\bibfnamefont {A.}~\bibnamefont
  {Matos-Abiague}}, \bibinfo {author} {\bibfnamefont {C.}~\bibnamefont
  {Ertler}}, \bibinfo {author} {\bibfnamefont {P.}~\bibnamefont {Stano}}, \
  and\ \bibinfo {author} {\bibfnamefont {I.}~\bibnamefont {Zutic}},\
  }\href@noop {} {\bibfield  {journal} {\bibinfo  {journal} {Acta Physica
  Slovaca}\ }\textbf {\bibinfo {volume} {57}},\ \bibinfo {pages} {342}
  (\bibinfo {year} {2007})}\BibitemShut {NoStop}%
\bibitem [{\citenamefont {Stoner}\ and\ \citenamefont
  {Wohlfarth}(1948)}]{Stoner1948}%
  \BibitemOpen
  \bibfield  {author} {\bibinfo {author} {\bibfnamefont {E.~C.}\ \bibnamefont
  {Stoner}}\ and\ \bibinfo {author} {\bibfnamefont {E.~P.}\ \bibnamefont
  {Wohlfarth}},\ }\href {\doibase 10.1098/rsta.1948.0007} {\bibfield  {journal}
  {\bibinfo  {journal} {Philosophical Transactions of the Royal Society A:
  Mathematical, Physical and Engineering Sciences}\ }\textbf {\bibinfo {volume}
  {240}},\ \bibinfo {pages} {599} (\bibinfo {year} {1948})}\BibitemShut
  {NoStop}%
\end{thebibliography}%

\newpage
\begin{center}
\textbf{\large Supplementary Information}
\end{center}
\setcounter{equation}{0}
\setcounter{figure}{0}
\setcounter{table}{0}
\setcounter{page}{1}

\renewcommand{\theequation}{S\arabic{equation}}
\renewcommand{\thefigure}{S\arabic{figure}}

\section{Fabrication details}
\begin{figure}[htb]
\centering
	\includegraphics[width=1\linewidth]{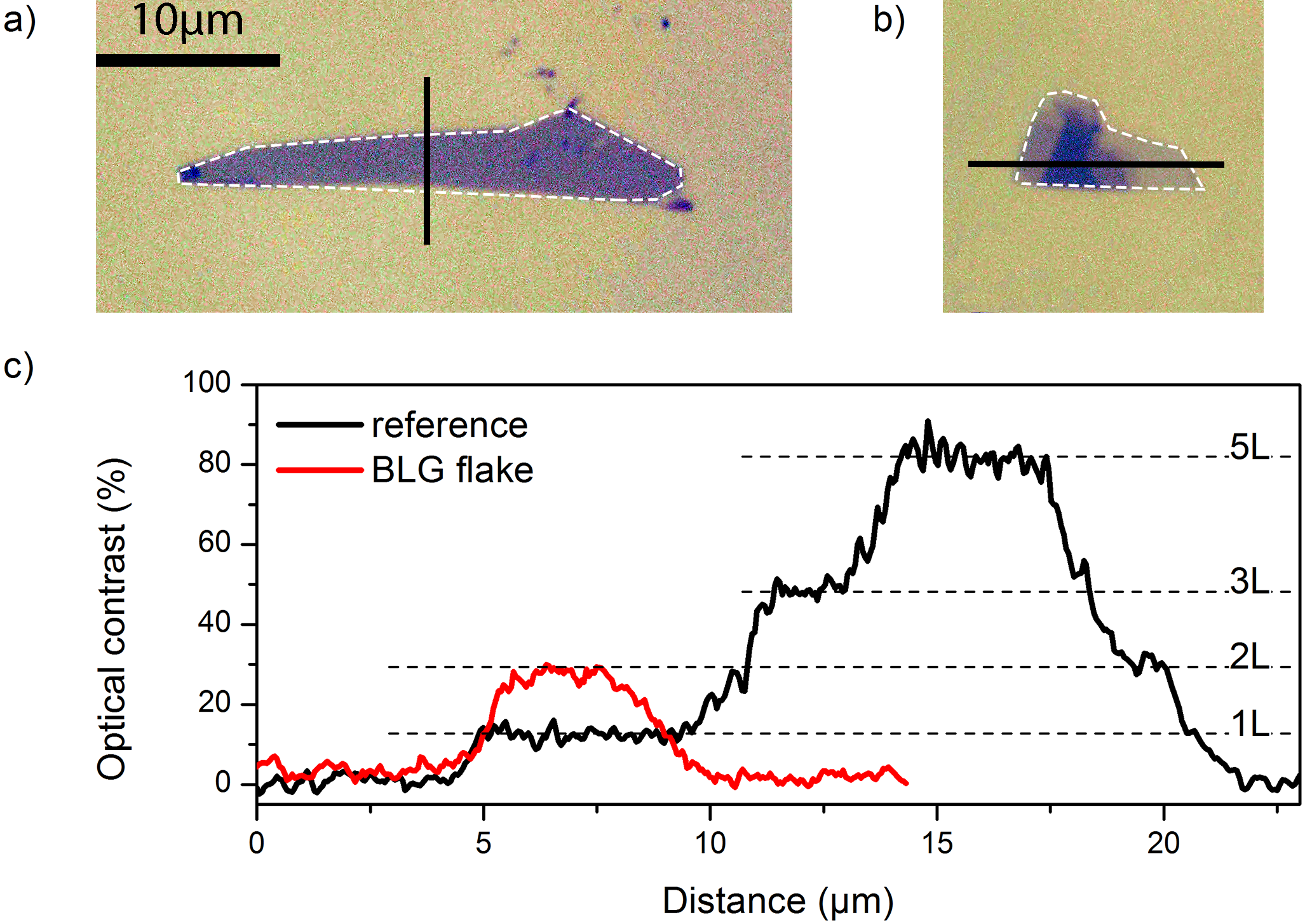}
	\caption{Optical image of the used BLG flake a) and the contrast reference flake b). The dashed white lines mark the edges of the flakes. The black line indicates the position where the optical contrast is measured. c) The contrast analysis confirms the graphene thickness to be two layers.
	}
	\label{SI_contrast}
\end{figure}

Thin hBN flakes are exfoliated from hBN powder (HQ Graphene) onto 90~nm SiO$_2$ wafers. Suitable hBN flakes are selected by their optical contrast and the thin-hBN/BLG/bottom-hBN stack is fabricated using a polycarbonate based dry transfer technique \cite{Zomer2014}. The bottom hBN flake has a thickness of 5~nm. The use of a thin-hBN flake ($\sim$ 1~nm, trilayer) as tunnel barrier for spin injection allows us to measure spin transport in a fully encapsulated high quality bilayer graphene device. Fig.~\ref{SI_contrast} shows the optical image and optical contrast analysis of the used BLG flake exfoliated from HOPG (HQ Graphene) on a 300~nm SiO$_2$ wafer. Its optical contrast, shown in Fig.~\ref{SI_contrast}c, is twice the single layer contrast, which is determined from the reference flake image in Fig.~\ref{SI_contrast}b. The BLG thickness is confirmed by atomic force microscopy and is $\sim$0.8~nm.

\begin{figure}[htb]
\centering
	\includegraphics[width=0.9\linewidth]{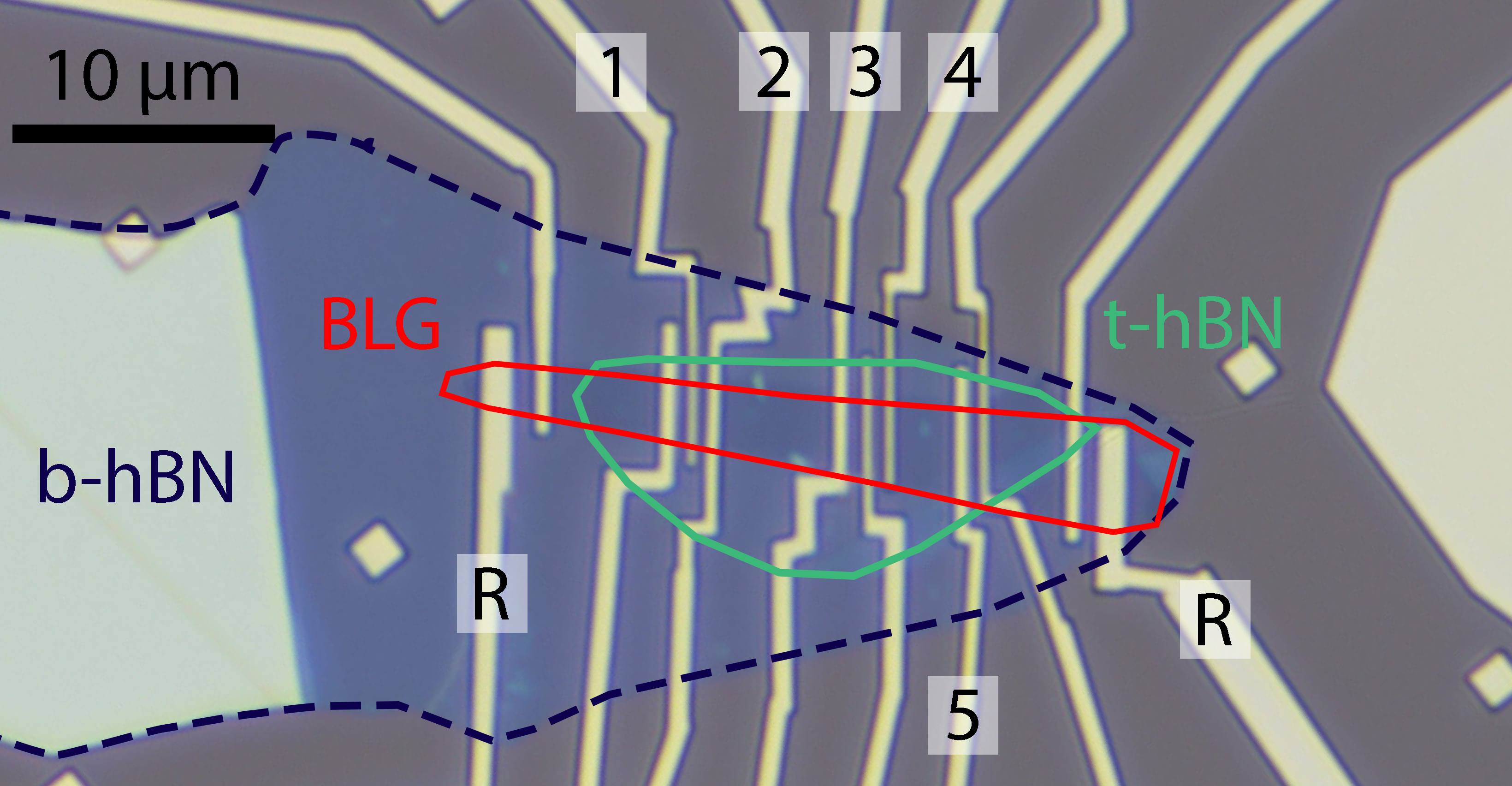}
	\caption{Optical image of the finished sample with labeled contacts. The outermost contacts are used as reference electrodes and do not have an hBN tunnel barrier.
	}
	\label{SI_optical}
\end{figure}

After the removal of the transfer polymer in chloroform the sample is annealed (1h in Ar/H$_2$ atmosphere) to clean the hBN surface and promote the adhesion of the metal film. Contacts are defined using standard two step PMMA-based e-beam lithography. Markers are exposed and developed in a first step and used for the contact exposure as reference. After development, the sample is loaded to an e-beam deposition system and 65~nm of cobalt are evaporated at a base pressure below 10$^{-7}$~mbar. Additionally, a 5~nm aluminum capping layer is deposited to prevent the oxidation of the cobalt. After liftoff in warm acetone, the finished device (Fig.~\ref{SI_optical}) is loaded into a cryostat where the sample space is evacuated below 10$^{-6}$~mbar.

\section{Charge and spin transport characterization}
\begin{figure}[htb]
\centering
	\includegraphics[width=0.9\linewidth]{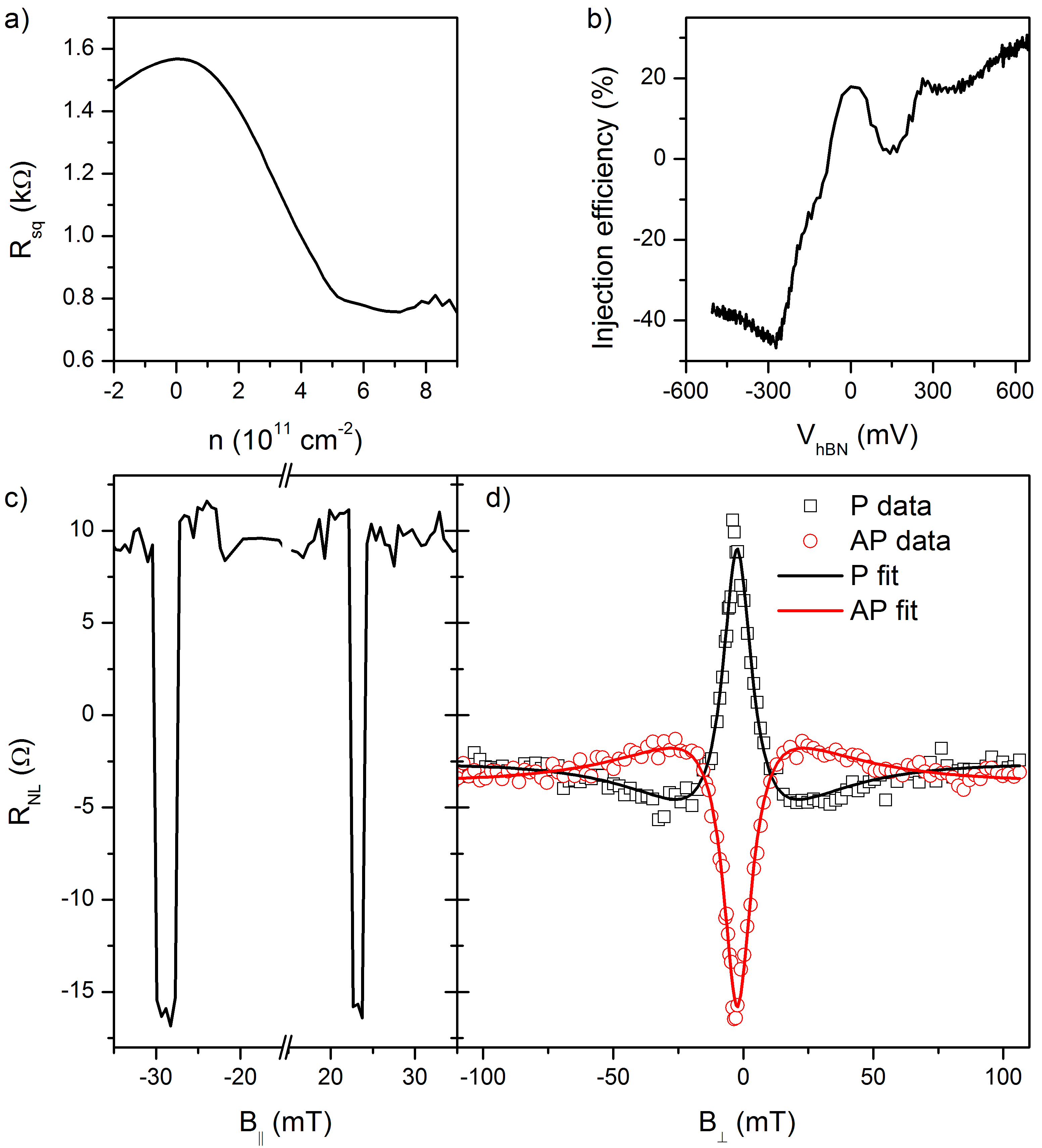}
	\caption{a) Dependence of the BLG square resistance on the carrier density n. b) DC bias dependence of the spin injection efficiency of contact 2 (injector used in the main text). c) Spin valve measurement of the device. d) Spin precession in (anti)parallel alignment of the injector and detector electrode.
	}
	\label{FigureSpinT_SI}
\end{figure}

The carrier density dependence of the square resistance $\rsq$ of the BLG flake between contacts 1 and 3 is shown in Fig.~\ref{FigureSpinT_SI}a). We can tune the carrier concentration n through the 90~nm SiO$_2$ and the 5~nm thick b-hBN from $6\times 10^{11}$~cm$^{-2}$ in the electron regime to slightly beyond the charge neutrality point (CNP) at $2\times 10^{11}$~cm$^{-2}$ in the hole regime. In this range we observe a gate leakage current below 10~nA. The carrier concentration in BLG is calculated via:
\begin{align}
\mathrm{n} = \epsilon_0 \epsilon (\mathrm{V}_\mathrm{BG}-\mathrm{V}_\mathrm{CNP})/(\mathrm{t}_\mathrm{BG} \cdot e)
\end{align}
where $\epsilon_0 = 8.854 \times 10^{-12}$~F/m denotes the vacuum permittivity, $\epsilon = 3.9$ the relative dielectric permittivity of SiO$_2$, $\mathrm{V}_\mathrm{BG}$ the voltage applied to the back gate, $\mathrm{V}_\mathrm{CNP}$ = -2~V the gate voltage at the CNP and t the thickness of the gate oxide. Here we assume that the dielectric permittivity of hBN has approximately the same value as SiO$_2$ and use the dielectric thickness of t$_\mathrm{BG}$ = t$_\mathrm{SiO_2}$ + t$_\mathrm{hBN}$ = 95~nm. Note that the gate leakage current increased during the measurements and prohibited in the end to reach the CNP at room temperature. 

The basic characterization of the spin transport in the non local geometry is shown in Fig.~\ref{FigureSpinT_SI}b and c. Here we use, as in the main text, contact~1 as injector and contact~4 as detector electrodes. The contacts are separated by L = 7~$\mu$m. We use the outermost contacts as reference electrodes which do not have a tunnel barrier.
We source an AC current of 50~nA between the ferromagnetic injector and the left reference electrode (R). A spin accumulation is injected through the hBN tunnel barrier and diffuses along the BLG flake. The detector probes the spin accumulation underneath its contact relative to the right reference electrode as $\mathrm{V}_\mathrm{NL}$.  In this particular measurement we do not apply any DC bias or gate voltage, n is here $4\times 10^{11}$~cm$^{-2}$ in the electron regime.

We observe a signal of $\rnl = \mathrm{V}_\mathrm{NL}/ \mathrm{I}_\mathrm{AC} = 25~\ohm$ in the spin valve, Fig.~\ref{FigureSpinT_SI}b. The spin precession in a perpendicular magnetic field $\bper$ in (anti)parallel alignement is shown in Fig.~\ref{FigureSpinT_SI}c.
By fitting the Hanle spin precession data we extract the spin relaxation time $\tsp$ = (1.9 $\pm$ 0.2)~ns and a spin diffusion constant $\dsp$ = (201 $\pm$ 32)~cm$^{2}$/s of our device and calculate the in-plane spin relaxation length $\lambda_\parallel = \sqrt{\dsp \tsp} \sim$ 6.2~$\mu$m.


From the measurements of the spin valve signals without any DC bias current in three different configurations with alternating injector and detector combinations we extract an unbiased spin polarization of 21\%, which is consistent throughout all measured contacts. A characteristic feature of spin injection from cobalt electrodes into graphene through hBN tunnel barriers is the dependence of the spin injection efficiency on the voltage applied across the hBN tunnel barrier. We found that a positive bias increases the spin injection efficiency and a negative bias also results in a sign change in the spin injection and consequently in the $\rnl$ \cite{Gurram2017,Gurram2018}. For the data shown in the main text we apply, additionally to the AC current, a DC bias current of -0.6~$\mu$A, which corresponds to a voltage of -300~mV and increases the unbiased spin injection efficiency from 21\% to above -40\%. The spin injection efficiency of the injector as a function of the applied DC bias is shown in Fig.~\ref{FigureSpinT_SI}b. The DC bias improves the signal to noise ratio which significantly enhances the data quality for measurements at the CNP. Note that the negative DC bias changes also the sign of $\rnl$. To avoid confusion with the conventional sign of $\rnl$, we have inverted the sign for all biased Hanle curves. Our analysis and the resulting claims are not affected by this.

\section{Estimation of the electric field}
To determine the electric field applied to the BLG flake we try to estimate the doping at the top and bottom side of the BLG. Since we have only one gate, we cannot control the electric field and carrier density independently. Hence, we estimate the lower bound of the electric field under the assumption that the doping is equal at both sides of the BLG flake. The carrier density is then determined by:
\begin{align}
\mathrm{n} = \epsilon_0 \epsilon \mathrm{V}_\mathrm{BG}/(\mathrm{t}_\mathrm{BG} \cdot e) + \mathrm{n}_\mathrm{bottom} + \mathrm{n}_\mathrm{top}
\end{align}
where $\mathrm{n}_\mathrm{top}$ and $\mathrm{n}_\mathrm{bottom}$ are the carrier densities induced by the doping at the top and bottom sides of the BLG flake. The external electric field is then defined as:
\begin{align}
\mathrm{\overline{E}} = \mathrm{V}_\mathrm{BG}/2\mathrm{t}_\mathrm{BG} - \mathrm{n}_\mathrm{bottom}/2\epsilon + \mathrm{n}_\mathrm{top}/2\epsilon
\end{align}
When assuming that $\mathrm{n}_\mathrm{bottom} = \mathrm{n}_\mathrm{top}$, we obtain as lower bound:
\begin{align}
\mathrm{\overline{E}}_\mathrm{CNP} = \mathrm{V}_\mathrm{BG}/2\mathrm{t}_\mathrm{BG} \sim 10\, \mathrm{mV/nm}
\end{align}
Assuming that all doping arises from the BLG top, $\mathrm{n}_\mathrm{bottom} = 0$, we obtain as upper bound $\mathrm{\overline{E}}_\mathrm{CNP} = 20$~mV/nm.

\section{Measurements using different injector-detector spacings}
\begin{figure}[htb]
\centering
	\includegraphics[width=1\linewidth]{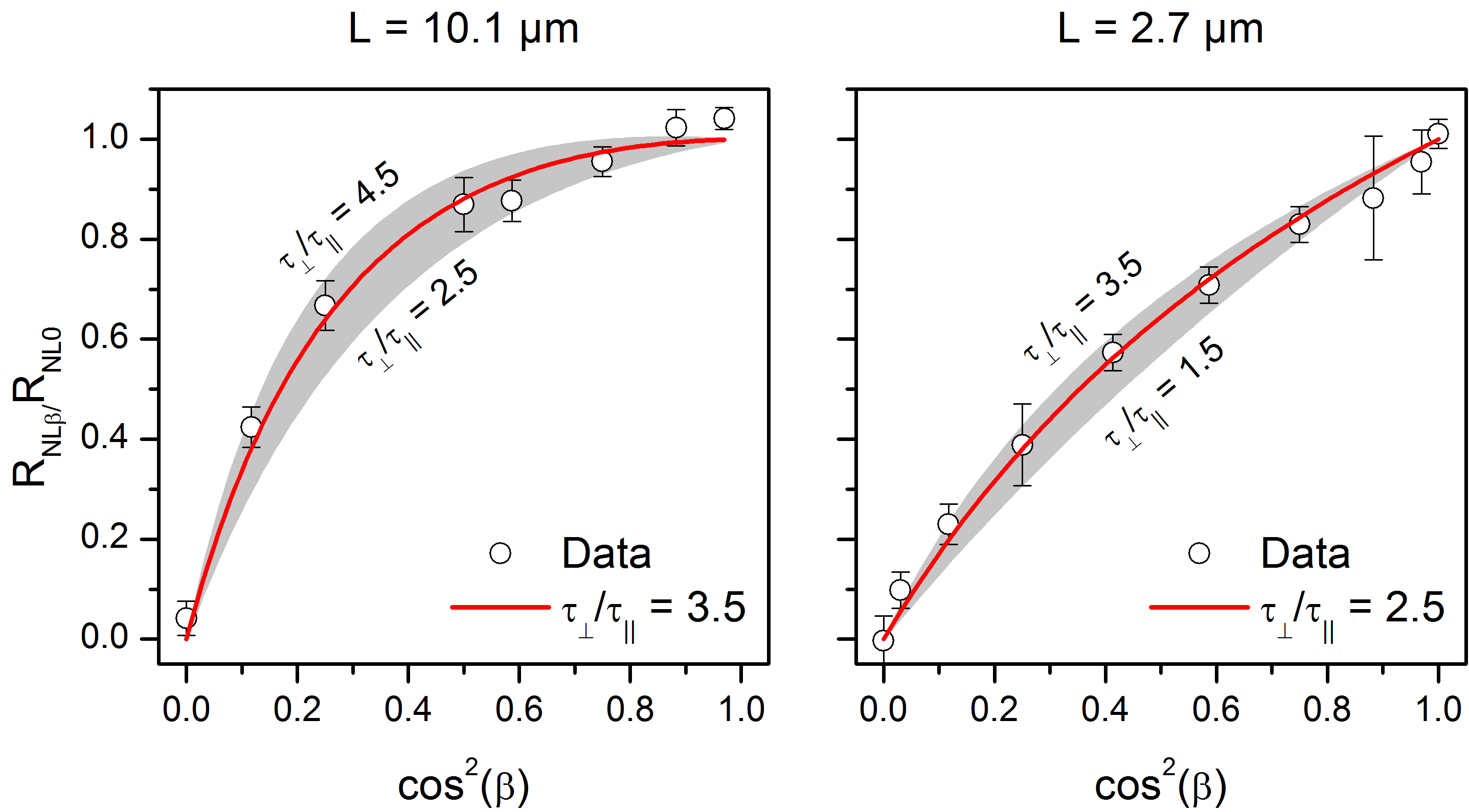}
	\caption{Extraction of the spin lifetime anisotropy for a) L = 10.1  $\mu$m (contacts 1 and 5) and b) L = 2.7 $\mu$m (contacts 4 and 5) at n = $6\times 10^{11}$~cm$^{-2}$ and T = 75~K. The shaded area corresponds to the estimated error margin.
	}
	\label{SI_length}
\end{figure}

Fig.~\ref{SI_length} contains the $\mathrm{R}_{\mathrm{NL}\beta}/\mathrm{R}_\mathrm{NL0}$ ratio for two different injector-detector spacings measured at T = 75~K and n = $6\times 10^{11}$~cm$^{-2}$. The measurements presented in the main text in Fig.~2a and d have yielded $\tper/\tpar$ = 3.5 for the same carrier concentration. 

Fig.~\ref{SI_length}a is measured at a longer spacing of L = 10.1~$\mu$m where contact 1 is used as injector and 5 as detector. Fig.~\ref{SI_length}b uses contact 4 as injector and 5 as detector, where L = 2.7~$\mu$m. 
For L = 10.1~$\mu$m, we find a similar value as discussed in the main text of $\tper/\tpar$ = 3.5 $\pm$ 1. With a different injector contact and a shorter spacing of L = 2.7~$\mu$m, we extract a slightly smaller value. 
Within the experimental uncertainty, all different spacings and injector and detector configurations yield similar anisotropies. As a consequence we conclude that our device is homogeneous and the results from our analysis do not depend on the specific contact pair used.

\section{Spin precession measurements with in-plane magnetic fields}
\begin{figure}[htb]
\centering
	\includegraphics[width=1\linewidth]{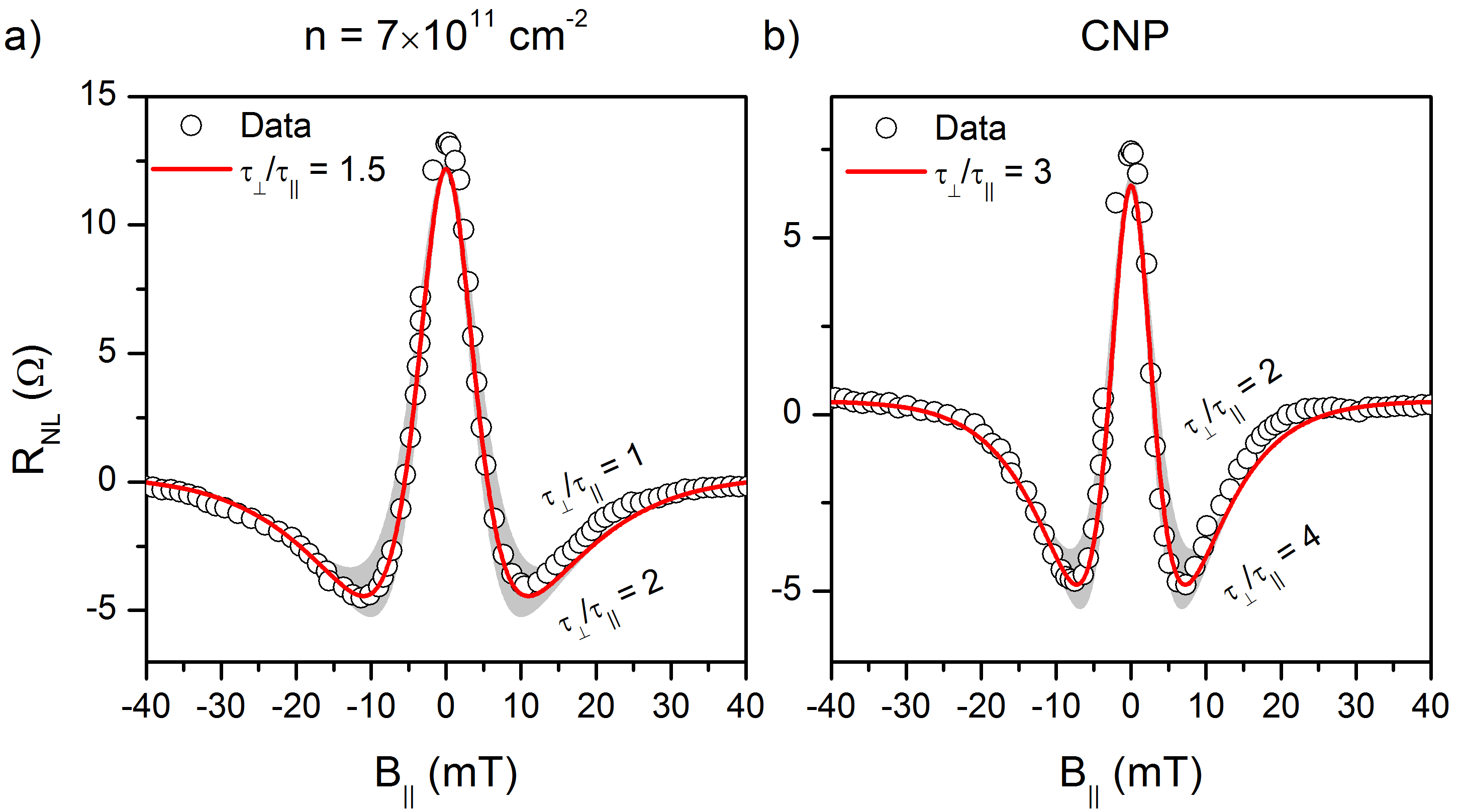}
	\caption{In-plane spin precession measurements over L = 10.1~$\mu$m at T = 75~K and two different carrier concentrations. The gray area corresponds to the estimated error margin, the red line to the fit of $\tper/\tpar$.}
	\label{SI_data_ip_Hanle}
\end{figure}

Fig.~\ref{SI_data_ip_Hanle} contains the measurements of the spin precession with an in-plane magnetic field perpendicular to the injected spin direction, along the device length. In this experiment the magnetic field rotates the injected spins in the $\bpar$ and $\bper$ plane. Therefore, both in-plane and out-of-plane spin lifetimes will be probed. 

The data shown in Fig.~\ref{SI_data_ip_Hanle} is measured with contact~1 as injector and 5 as detector, L = 10.1~$\mu$m. $\rnl$ is extracted from the spin precession measurement in (anti)parallel electrode configuration. 

Using the model described below that accounts for the actual device geometry we model anisotropic spin transport and estimate $\tper/\tpar \sim 1.5$ at $6\times 10^{11}$~cm$^{-2}$ and $\tper/\tpar \sim 3$ near the CNP. In comparison to the oblique spin precession measurements we find slightly smaller anisotropies, which is consistent with Ref.~\cite{Ringer2017}. 
We attribute this observation to a change in the sample parameters that occurred prior to this measurement due to unloading of the sample from the cryostat. Nevertheless, the anisotropy remains tunable with the applied gate voltage.

\section{Low temperature anisotropy measurements}
\begin{figure}[htb]
\centerline{\includegraphics[width=1\linewidth]{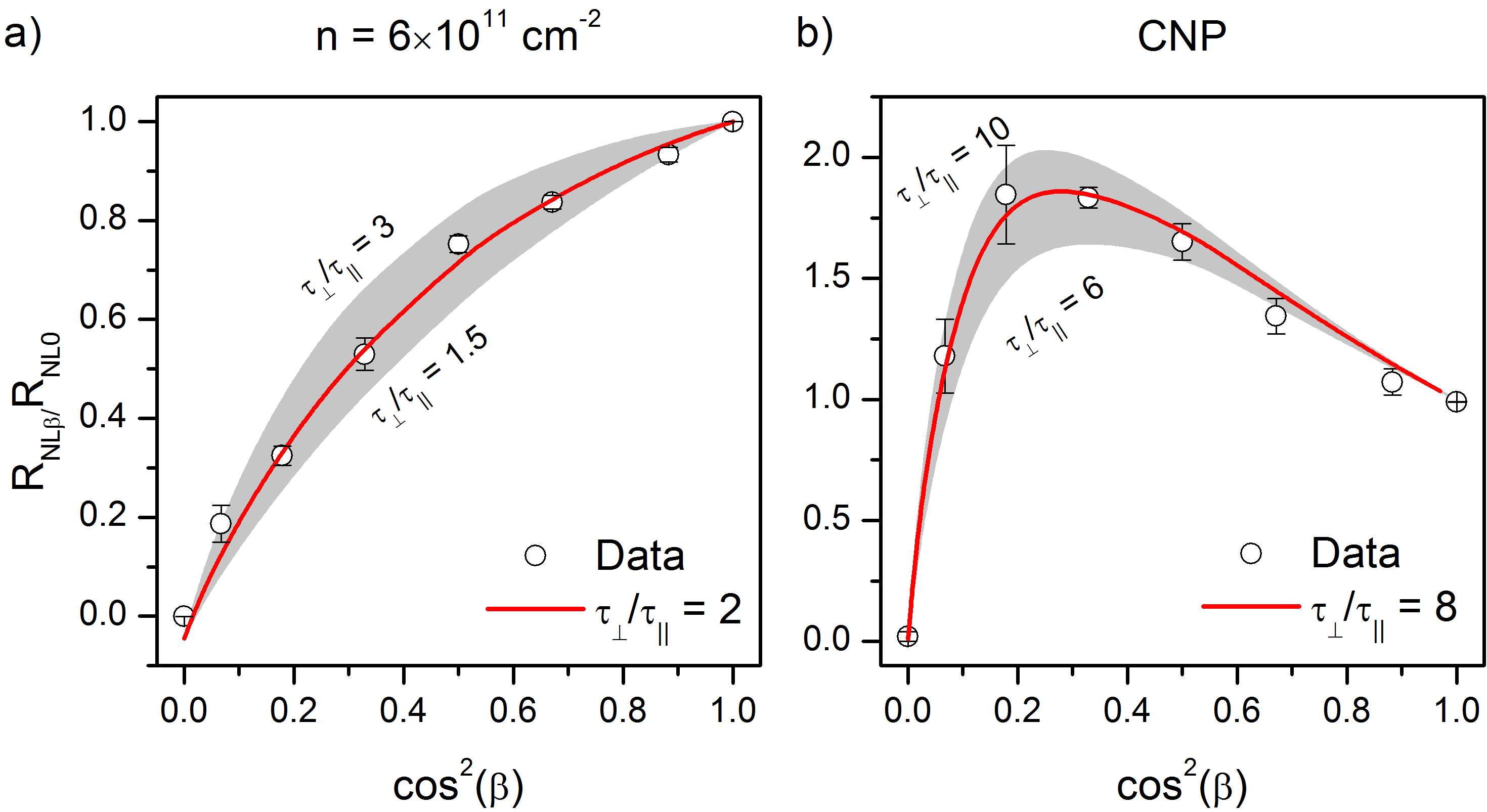}}
\caption{$\mathrm{R}_{\mathrm{NL}\beta}/\mathrm{R}_\mathrm{NL0}$ and the extracted $\tper/\tpar$ ratio at T = 5~K gives similar anisotropies as the measurements at T = 75~K discussed in the main text.
\label{5K_data_SI}}
\end{figure}

Fig.~\ref{5K_data_SI} contains the $\mathrm{R}_{\mathrm{NL}\beta}/\mathrm{R}_\mathrm{NL0}$ ratio extracted from oblique Hanle measurements at T = 5~K using contacts 1 and 3 as injector and detector (L = 5.2~$\mu$m). In comparison to the measurements at 75~K and L = 7~$\mu$m, we find a very comparable values of the spin lifetime anisotropy and dependence on the carrier density. 

\section{Carrier concentration dependence of the in-plane spin lifetime}
\begin{figure}[htb]
\centerline{\includegraphics[width=0.8\linewidth]{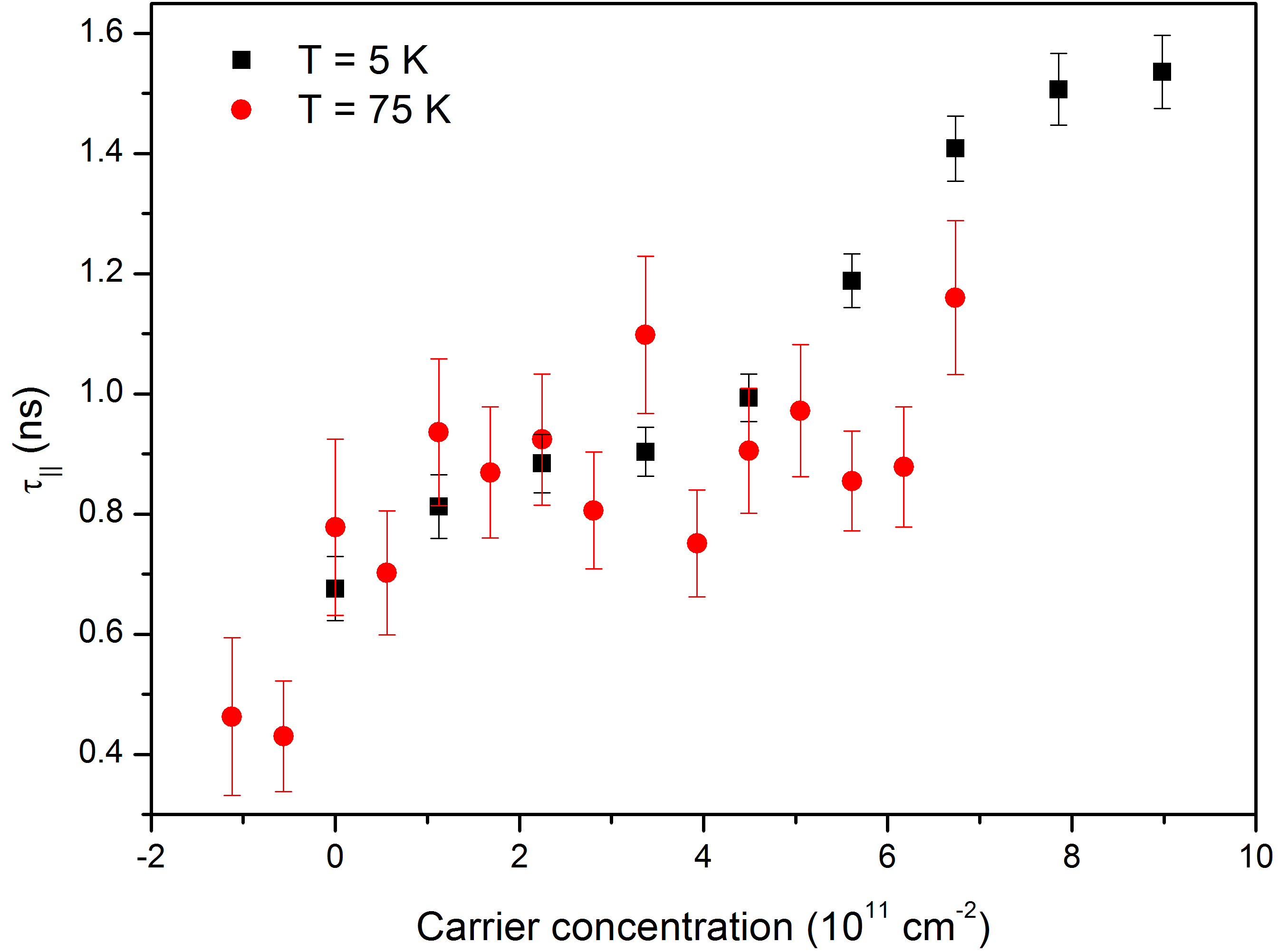}}
\caption{Carrier density dependence of the in-plane spin lifetime at 5~K (black squares) and 75~K (red circles) measured with $\bper$.
\label{SI_ipSLA}}
\end{figure}

We have measured the carrier density dependence of the in-plane spin lifetime at 5~K and 75~K (Fig.~\ref{SI_ipSLA}). As a result we obtained that, at both temperatures, $\tpar$ increases with increasing density in the conduction band. This result is in contrast with other reports of bilayer graphene on SiO$_2$ \cite{Han2011,Yang2011,Avsar2011}, where the opposite trend was observed at 5~K.

\section{Carrier density dependence of the magnetoresistance}
\begin{figure}[htb]
\centerline{\includegraphics[width=0.8\linewidth]{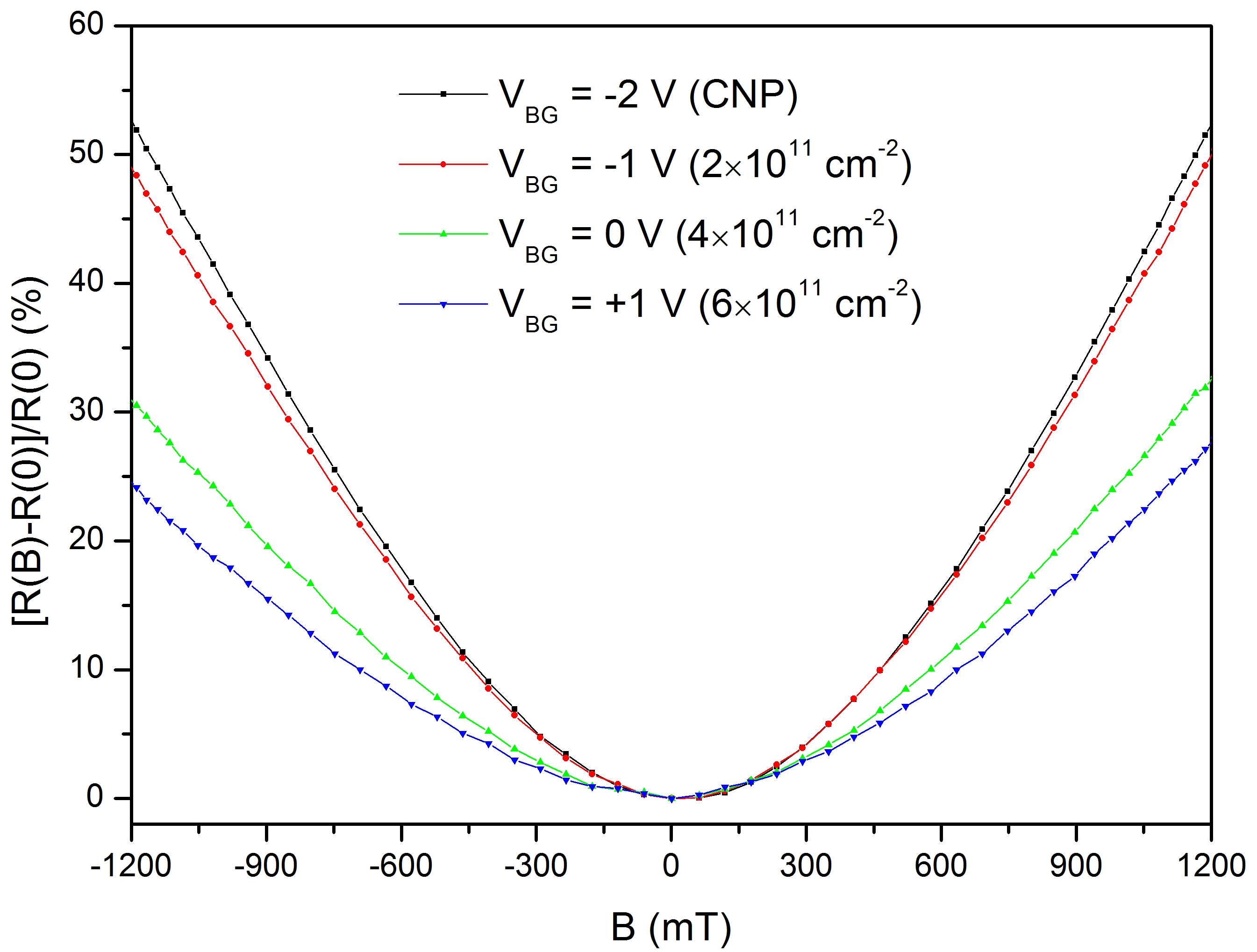}}
\caption{Magnetoresistance of the graphene channel at different gate voltages at T = 75~K.
\label{FigureSI_MR}}
\end{figure}

Fig.~\ref{FigureSI_MR} shows the four probe magnetoresistance of the graphene channel. The magnetoresistance is negligible and less than 1\% at low magnetic fields between 50 and 100~mT. Therefore, it does not affect our low field analysis. At higher magnetic fields of 1.2~T, the magnetoresistance reaches up to 50\% at the CNP. At higher carrier densities this value decreases to 25\%. Since the possible contribution of magnetoresistance to $\rnl$ depends on the background resistances which are smaller than 20~$\ohm$ and the agreement between the low field and high field analysis, we conclude that the effect is not dominant for the high field analysis. 

\section{Modeling of the spin lifetime anisotropy}\label{SectionModel}
As described in the main text, our device length is comparable to the in- and out-of-plane spin relaxation lengths. As a consequence, we have to take the effect of the finite length on the extracted parameters into account. Therefore, we use a numerical model that accounts for the following:
\begin{enumerate}
\item The spin lifetime anisotropy in the channel.
\item The finite length of the channel.
\item The effect of spin absorption by the reference contacts that do not have any tunnel barrier.
\item The effect of the magnetic field in the contact magnetization direction, which we estimate to have a maximum angle of 4$^\circ$ for $\beta=90^\circ$ at $\mathrm{B=0.1}$~T.
\end{enumerate}

\begin{figure}[htb]
\centerline{\includegraphics[width=0.8\linewidth]{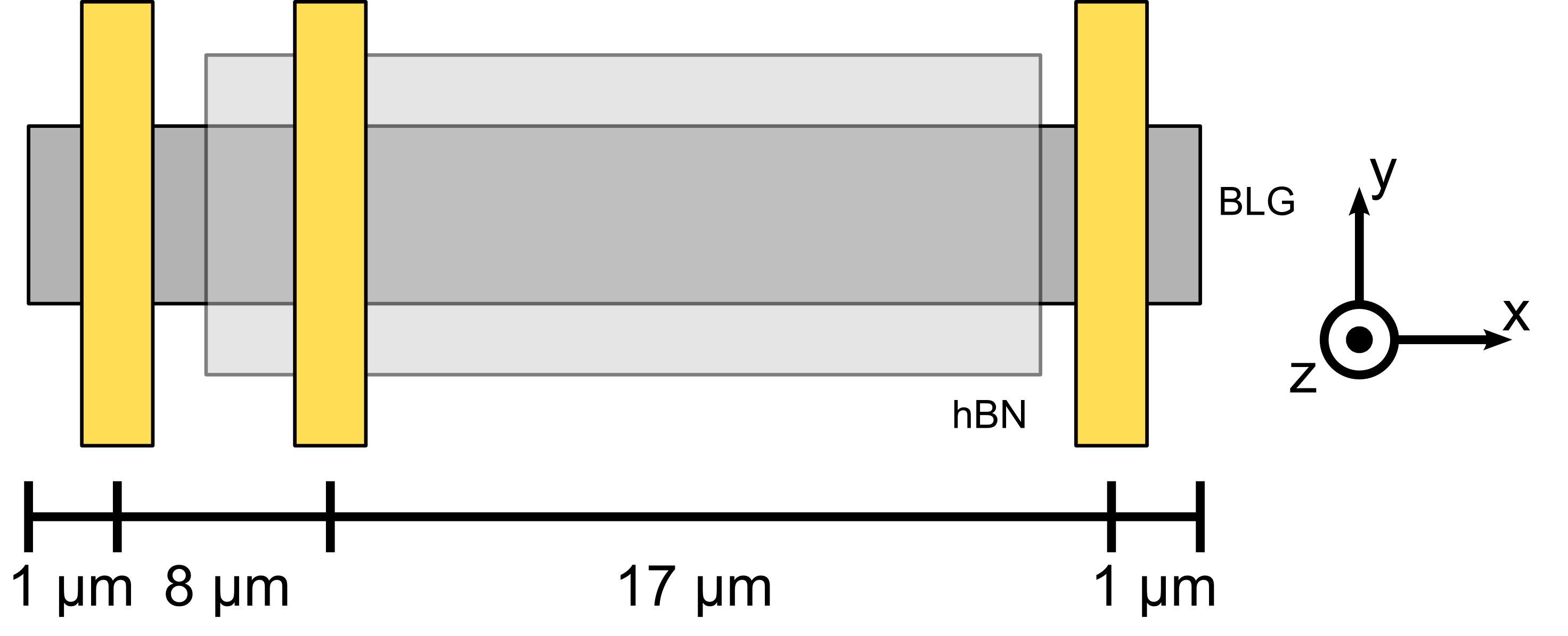}}
\caption{Sketch of the simulated device geometry.
\label{FigureSI_sketch}}
\end{figure}

The model is based on the Bloch equations with anisotropic spin relaxation \cite{Fabian2007,Raes2016} using the device parameters shown in table~\ref{table2SI} and geometry sketched in Fig.~\ref{FigureSI_sketch}.

\begin{align}
0 & = \ds\frac{d^2\mu_{\mathrm{sx}}}{d\mathrm{x}^2}-\frac{\mu_{\mathrm{sx}}}{\tpar}+\gamma \by\mu_{\mathrm{sz}}-\gamma \bz\mu_{\mathrm{sy}}\\
0 & = \ds\frac{d^2\mu_{\mathrm{sy}}}{d\mathrm{x}^2}-\frac{\mu_{\mathrm{sy}}}{\tpar}+\gamma \bz\mu_{\mathrm{sx}}-\gamma \bx\mu_{\mathrm{sz}}\\
0 & = \ds\frac{d^2\mu_{\mathrm{sz}}}{d\mathrm{x}^2}-\frac{\mu_{\mathrm{sz}}}{\tper}+\gamma \bx\mu_{\mathrm{sy}}-\gamma \by\mu_{\mathrm{sx}}
\label{BlochSup}
\end{align}
where $\vec{\mu}_\mathrm{s}=(\mu_{\mathrm{sx}},\mu_{\mathrm{sy}},\mu_{\mathrm{sz}})$ is the three dimensional spin accumulation, $\ds$ is the spin diffusion coefficient, $\tpar$ and $\tper$ are the in- and out-of-plane spin relaxation times, and $\mathrm{\gamma \mathrm{B}=g\mu_\mathrm{B} \mathrm{B}/\hbar}$ is the Larmor frequency with the Land\'e factor $g = 2$, $\mathrm{\mu_B}$ the Bohr magneton and $\hbar$ the reduced Planck constant. In our devices, the ferromagnetic contacts go all across the channel. This makes the spin accumulation constant over the sample width ($\mathrm{W_s}$) and allows us to make our analysis 1D. Here we use the average width ($3~\mu$m) of the relevant region of the BLG flake.

\begin{table}[htb]
\caption{Device parameters used in the model. $\mathrm{L_{l}}$ denotes the distance from the injector to the left sample edge, $\mathrm{L_{r}}$ the distance to right sample edge, $\mathrm{L_{cl}}$ the distance to the injector reference with the contact resistance $\mathrm{R_{cl}}$ and $\mathrm{L_{cr}}$ the distance to the voltage reference with $\mathrm{R_{cr}}$.}
        \begin{ruledtabular}
        \begin{tabular}{c c c c c c}
            \renewcommand{\arraystretch}{2}
            $\mathrm{L_{l}}$~($\mu$m) & $\mathrm{L_{r}}$~($\mu$m)& $\mathrm{L_{cl}}$~($\mu$m) & $\mathrm{L_{cr}}
            $~($\mu$m) & $\mathrm{R_{cl}}$~($\Omega$) & $\mathrm{R_{cr}}$~($\Omega$) \\
            \hline
             8 & 18& 7 & 17 & 500&500\\

    \end{tabular}
    \end{ruledtabular}
    \label{table2SI}
\end{table}

The magnetization direction is determined using the Stoner-Wohlfarth model \cite{Stoner1948}. Because the magnetic field is applied in the y-z plane, we solve the Stoner-Wohlfarth equation numerically:
\begin{align}
\mathrm{\sin(2(\phi-\beta))/2+h\sin(\phi)=0} 
\end{align}
Where $\mathrm{h=B/B_{sat}}$ is the effective external field. $\mathrm{B_{sat}}$ is the field at which the electrode magnetization saturates in the direction perpendicular to the easy axis. In our case, we assume that $\mathrm{B_{sat}}=1.5$~T based on earlier measurements of comparable cobalt electrodes with similar thickness.
As defined in the main text, $\beta$ is the angle between the magnetic field and the easy axis of the ferromagnet, $\phi$ is the angle between the contact magnetization and the applied magnetic field. The angle between the magnetization M and the easy axis is $\gamma=\beta-\phi$. 
To determine the spin signal in the channel we use the following boundary conditions:
\begin{itemize}
\item The spin accumulation $\mu_\mathrm{s}$ is continuous everywhere.
\item The spin current is defined as $\mathrm{I_s=W_s/(2eR_{sq})(d\mu_{sx}/dx,d\mu_{sy}/dx,d\mu_{sz}/dx)}$ where $\mathrm{W_{s}}$ is the width of the graphene, $\mathrm{R_{sq}}$ is the square resistance of the graphene channel and $\mathrm{e}$ is the electron charge.
\item The spin current has a discontinuity of $\mathrm{\Delta I_s=I\cdot P_{inj}/2(0,\cos(\gamma),\sin(\gamma))}$ at the injection point.
\item The spin current is discontinuous at the transparent outer contacts due to the spin backflow effect. This discontinuity is of  $\mathrm{\Delta I_s= -I_{back}= -\mu_s/(2eR_c)(1,1,1)}$ where $\mathrm{R_c}$ is the resistance of the reference contacts.
\item The spin current at the sample end is zero. 
\end{itemize}

Using these equations, we have performed a finite difference calculation that implements an implicit Runge-Kutta method in Matlab to determine the spin signal.

\section{Effect of the contact resistance on the anisotropy}
The interface resistances of the outer contacts are comparable to the resistances of the cobalt leads. Therefore, it is not possible to determine their exact interface resistance from three terminal measurements. To estimate the resulting uncertainty, we have performed simulations of angle dependent spin precession with different contact resistances using the model described in the previous section. Here we use the spin transport parameters measured at n = $6\times 10^{11}$~cm$^{-2}$ and an anisotropy of $\tper/\tpar = 2.5$.

\begin{figure}[htb]
\centering
	\includegraphics[width=1\linewidth]{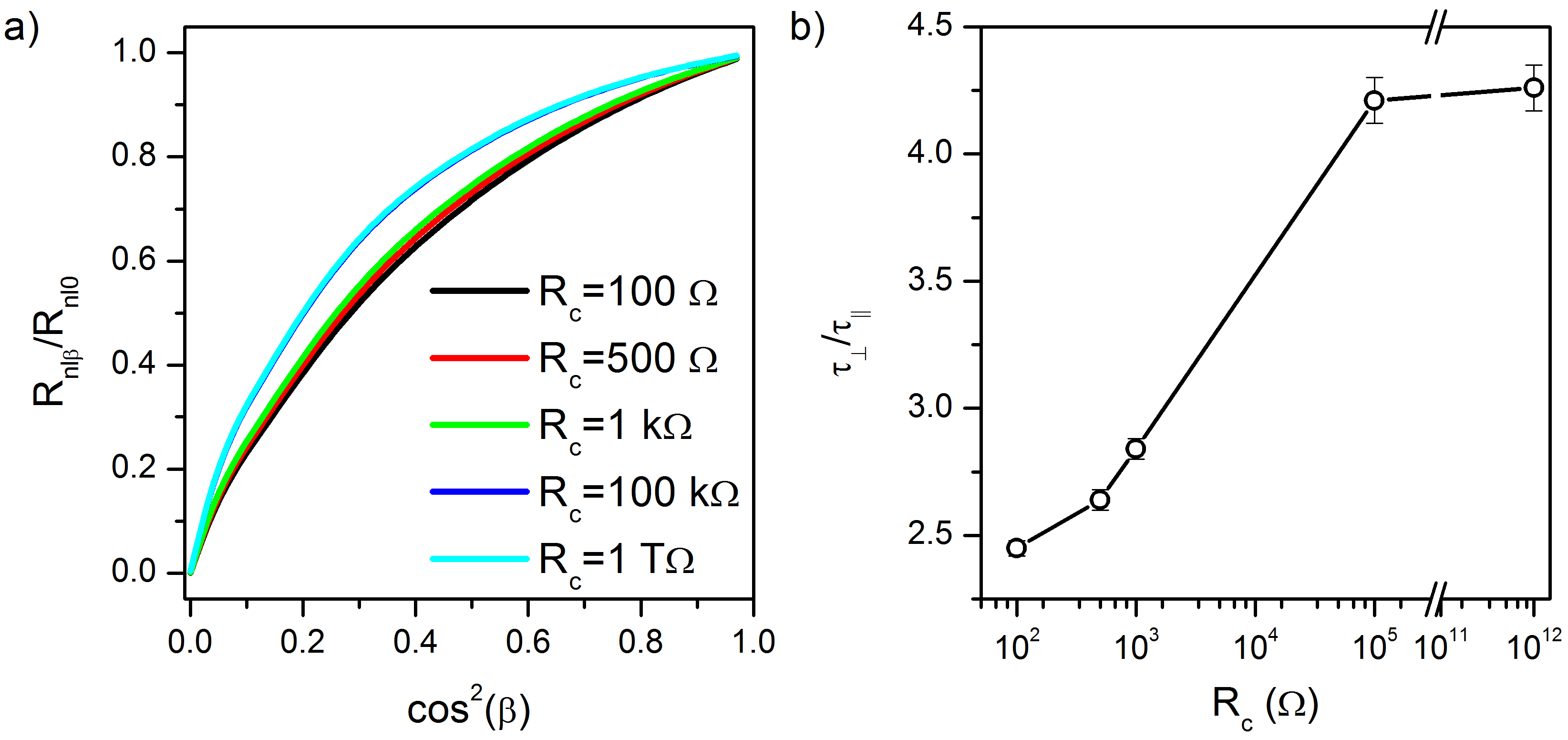}
	\caption{Effect of the contact resistance of the reference contacts on the ratio $\mathrm{R}_{\mathrm{NL}\beta}/\mathrm{R}_{\mathrm{NL}0}$ as a function of the angle $\beta$ between the B field and the y-axis a). b) The values of $\tper/\tpar$ are obtained from fits to Eq.~\ref{IdealModel1} for different contact resistances $\mathrm{R_c}$. The simulated anisotropy is $\tper/\tpar$ = 2.5 and is substantially overestimated by Eq.~\ref{IdealModel1}.
	}
	\label{EffectRc}
\end{figure}

The simulated Hanles are analyzed by evaluating the average signal between $\mathrm{B=0.05}$~T and $\mathrm{B=0.1}$~T. The output of this operation is defined as $\mathrm{R_{NL \beta}}$ and is normalized to the value of $\mathrm{R_{NL 0}}$ at B = 0 to obtain the ratio $\mathrm{R}_{\mathrm{NL}\beta}$/$\mathrm{R}_{\mathrm{NL}0}$. The angle dependence of $\mathrm{R}_{\mathrm{NL}\beta}$/$\mathrm{R}_{\mathrm{NL}0}$ is shown in Fig.~\ref{EffectRc}a for different contact resistances. To determine the effect of these changes in the spin lifetime anisotropy, we fit the results from a to the infinitely long channel model \cite{Raes2016}:

\begin{align}
\frac{\mathrm{R}_{\mathrm{NL}\beta}}{\mathrm{R}_{\mathrm{NL}0}} &= \sqrt{\frac{\tbet}{\tpar}} \exp \left(-\frac{\mathrm{L}}{\lambda_\parallel}\left(\sqrt{\frac{\tpar}{\tbet}}-1\right)\right)\cos^2(\beta) \label{IdealModel1}\\
\frac{\tbet}{\tpar} &=\left(\cos^2(\beta) +\frac{\tpar}{\tper} \sin^2(\beta) \right)^{-1}
\label{IdealModel2}
\end{align}

The results from this calculation are shown in Fig.~\ref{EffectRc}b. From those results we conclude that 
\begin{enumerate}
\item The finite device size, without the presence of invasive contacts, leads to a substantial overestimation of the lifetime anisotropy when using Eq.~\ref{IdealModel1}.
\item The anisotropy extracted from $\mathrm{R_c}$ = 100~k$\ohm$ is almost exactly the same as the high resistance reference ($\mathrm{R_c}$ = 1~T$\ohm$). As a consequence, the effect of the contact backflow when $\mathrm{R_c\geq 100}$~k$\Omega$ is negligible, which is the case for all contacts with an hBN tunnel barrier. Furthermore, this justifies that we do not have to take additional contacts between injector and detector electrodes into account.
\item The invasive reference contacts reduce the effect of the lifetime anisotropy on the measured signal, compensating for the confinement effect.
Since for those contacts $\mathrm{R_c}$ is lower than 500~$\Omega$, the absolute uncertainty in the anisotropy is about 0.25 and lower than the uncertainty in fitting the experimental data.
\end{enumerate}

\begin{table}[htb]
\caption{Spin and charge transport parameters used to determine the spin lifetime anisotropy in the bilayer graphene channel. The in-plane spin lifetime, spin diffusion coefficient and contact polarization are determined from Hanle precession with B applied perpendicular to the graphene plane. $\mathrm{R_{sq}}$ is obtained from local four probe measurements.}
        \begin{ruledtabular}
        \begin{tabular}{c c c c c}
            \renewcommand{\arraystretch}{2}
            $\mathrm{V_{bg}}$ (V) & $\mathrm{R_{sq}}$~($\Omega$)& $\ds$~($\mathrm{m^2/s}$) & $\tpar$~(ns) & P  \\
            \hline
             -2 & 1550& 0.010 & 1.1 & 0.42\\
             0 & 900& 0.018 & 1.87 & 0.431\\
             1 & 750& 0.021 & 1.74 & 0.472\\
    \end{tabular}
    \end{ruledtabular}
    \label{table1SI}
\end{table}

\section{Measurements on a second BLG device}
Lastly, we discuss the spin precession measurements of a BLG flake deposited on an Yttrium-Iron-Garnet (YIG) substrate. In contrast to our previous study of SLG on YIG, where we found an exchange field of the order of 0.2~T \cite{Leutenantsmeyer2017}, the exchange field of this BLG sample was determined to be below 4~mT and can therefore be neglected in the following analysis. 
This sample is not fully hBN encapsulated, only a bilayer hBN tunnel barrier is used for spin injection. Compared to the fully encapsulated sample, we observe significantly reduced spin lifetime, $\tpar$ = (99.1 $\pm$ 7.5)~ps and $\ds$ = (532 $\pm$ 41) cm$^2$/s. The in-plane spin relaxation length is 2.3 $\mu$m. The carrier concentration can not be directly measured in this type of samples. Similarly fabricated Hall bars show n $\sim$ $4\times 10^{12}$~cm$^{-2}$ and we expect the carrier concentration to be in a comparable range. 

\begin{figure}[H]
\centering
	\includegraphics[width=1\linewidth]{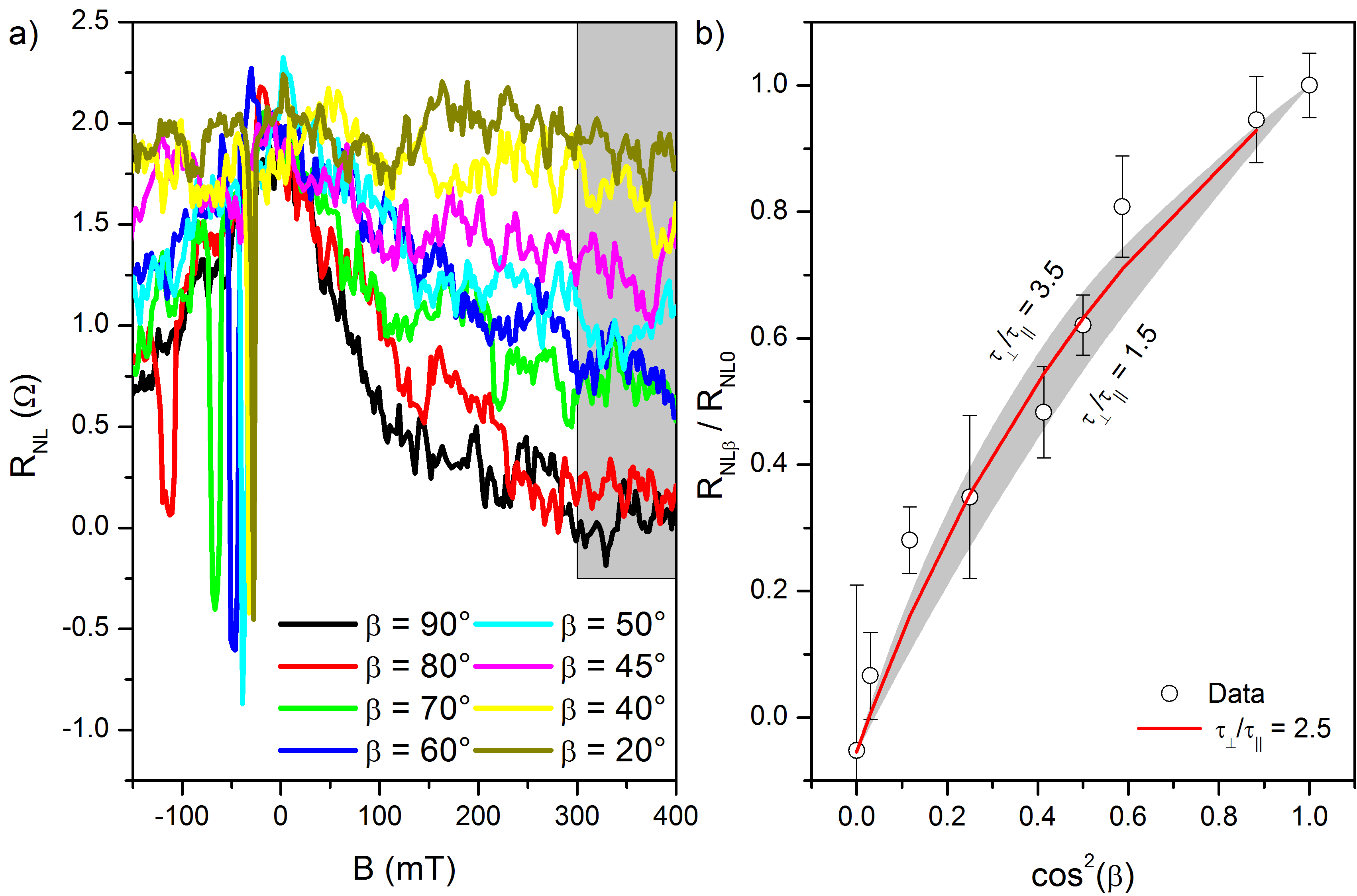}
	\caption{a) Measurements of the oblique Hanle spin precession in BLG without a bottom hBN flake. The reduced spin transport parameters require a larger field scan at which the in-plane field component starts to switch the injector and detector electrodes. b) The extracted $\mathrm{R}_{\mathrm{NL}\beta}/\mathrm{R}_\mathrm{NL0}$ ratio indicates $\tper/\tpar$ = 3.5. The data is measured at 75~K and an additionally applied DC bias current of -10~$\mu$A.
	}
	\label{Fig_SI_sample2}
\end{figure}

Fig.~\ref{Fig_SI_sample2} contains the $\mathrm{R}_{\mathrm{NL}\beta}/\mathrm{R}_\mathrm{NL0}$ ratio measured at different angles $\beta$. Note that the short values of $\tau$ cause a broadening of the Hanle curves. Therefore, we have to average the $\rnl$ at higher fields to obtain $\mathrm{R}_{\mathrm{NL}\beta}$ (300-400~mT). Nevertheless, we observe clearly anisotropic spin transport in the BLG flake, and $\mathrm{R}_{\mathrm{NL}\beta}$ at $\beta = 45^\circ$ is clearly above 0.5.

Fig.~\ref{Fig_SI_sample2}b shows the full analysis of the angle sweep. We extract $\tper/\tpar$ = 2.5 using our model. In comparison to the fully encapsulated BLG sample, we observe in this hBN-covered sample a smaller anisotropy, which we attribute to the difference in the carrier concentration of both samples. At $6\times 10^{11}$~cm$^{-2}$, we measured in the fully encapsulated device $\tper/\tpar$ = 3.5. An anisotropy value of $\tper/\tpar$ = 2.5 at around $4\times 10^{12}$~cm$^{-2}$ is therefore in good agreement with the carrier concentration dependence of the sample discussed in the main text. 

\end{document}